\documentclass[epj]{svjour}
\usepackage[T1]{fontenc}
\usepackage[latin1]{inputenc}
\usepackage{graphicx}

\makeatletter

\usepackage{color}
\usepackage{multibox}

\def\tensor#1{{\vec{#1}}}
\def\ellperso{\vec{\ell}}
\def\ellperso{\mbox{\boldmath $\ell$}}

\def\tensorstrain{\tensor{U}}

\def\eps{\mbox{\boldmath $\varepsilon$}}
\def\deps{{\dot{\mbox{\boldmath $\varepsilon$}}}}

\makeatletter
\usepackage{umlaut}
\makeatother
\begin{document}
\title{Discrete rearranging disordered patterns, part I: Robust statistical tools in two or three dimensions}
\titlerunning{Discrete rearranging patterns: Robust tools}
\author{F. Graner
\thanks{Author for correspondence at graner@ujf-grenoble.fr}
\and B. Dollet
\thanks{Present address: G.M.C.M. Universit\'{e} Rennes 1,  UMR CNRS 6626, B\^{a}timent 11A, Campus Beaulieu, 35042 Rennes Cedex, France}
\and C. Raufaste
\and P. Marmottant}
\institute{Laboratoire de Spectrom\'{e}trie Physique, UMR5588, CNRS-Universit\'{e} Grenoble I, B.P. 87, F-38402 St Martin
d'H\`{e}res Cedex, France}
\date{\today}
\abstract{
Discrete rearranging patterns include cellular patterns, for instance liquid foams, biological tissues, grains in polycrystals; assemblies of particles such as beads, granular materials, colloids, molecules, atoms; and interconnected networks.
Such a pattern can be described as a list of links between neighbouring sites.
Performing statistics on the links between neighbouring sites yields average quantities (hereafter "tools")
as the result of direct measurements on images.
These descriptive tools are flexible and suitable for various problems where quantitative measurements are required, whether in two or in three dimensions.
Here, we present a coherent set of robust tools, in three steps. First, we revisit the definitions of three existing tools based on the texture matrix. Second, thanks to their more general definition, we embed these three tools in a self-consistent formalism, which includes three additional ones. Third, we show that the six tools together provide a direct correspondence between a small scale, where they quantify the discrete pattern's local distortion and rearrangements, and a large scale, where they help describe a material as a continuous medium.
This enables to formulate elastic, plastic, fluid behaviours in a common, self-consistent modelling using continuous mechanics. Experiments, simulations and models can be expressed in the same language and directly compared.
As an example, a companion paper  \cite{Cartes2D}
 provides an application to foam plasticity.
} 
\maketitle

\section{Introduction}
\label{introduction}

\begin{figure}
(a) \includegraphics[width=6cm]{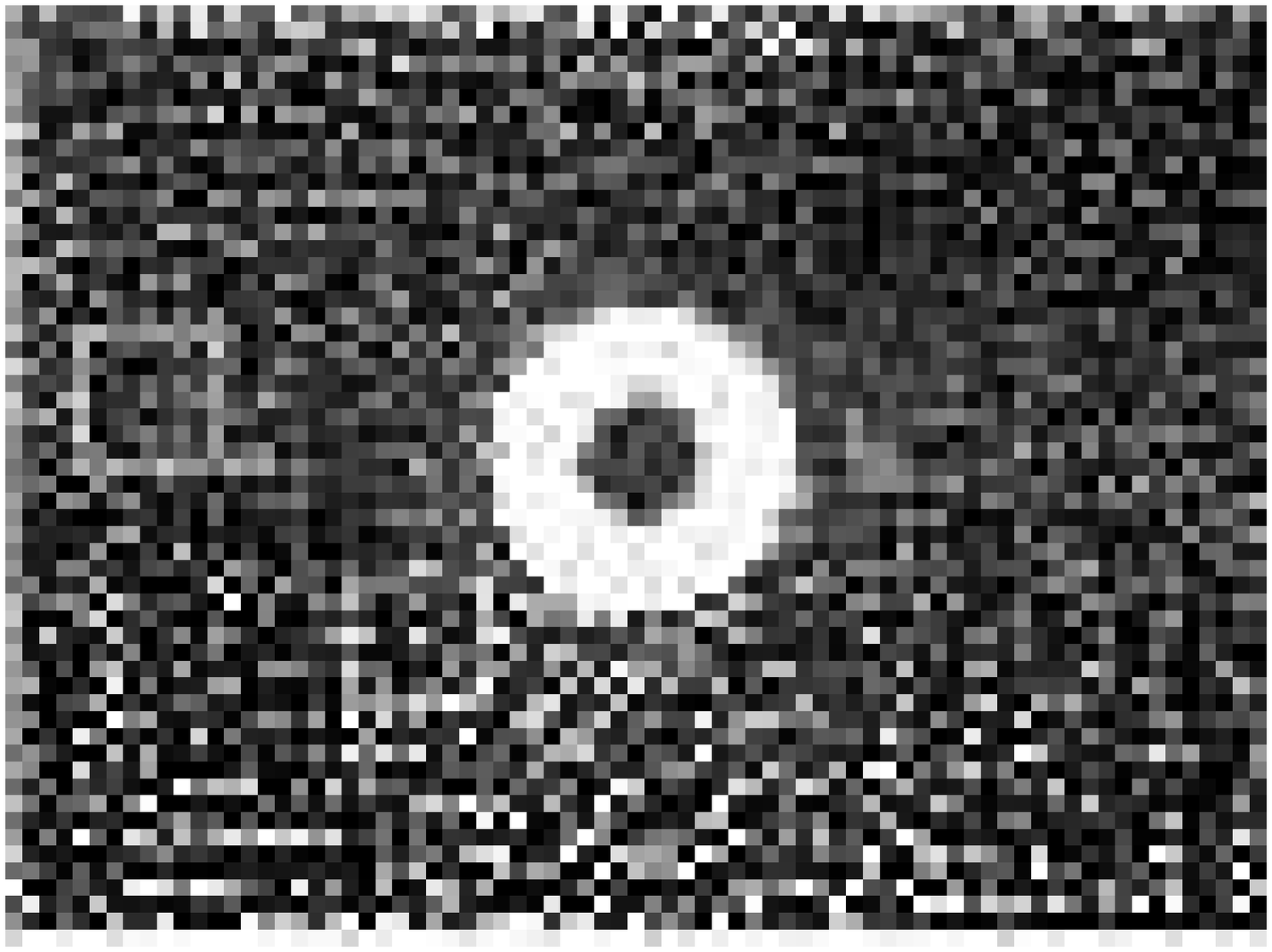}\\
(b) \includegraphics[width=6cm]{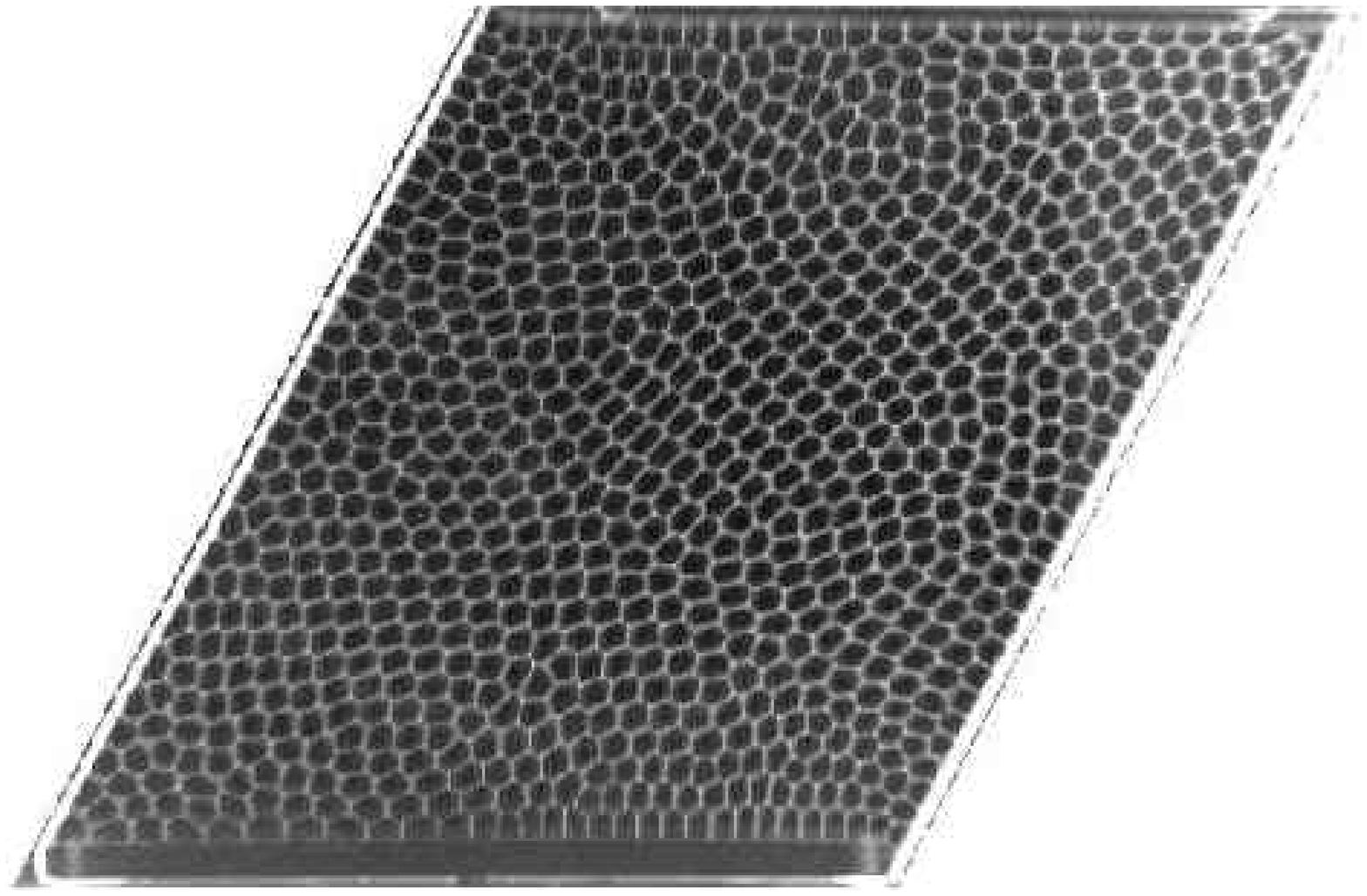}
\caption{
 Liquid foams. (a)
Heterogeneous flow: from left to right, around an obstacle \cite{rau07};
liquid fraction $\sim 10^{-4}$, image width: 15 cm.
(b)
Homogeneous shear: in a rectangular box, deformed at constant area \cite{Quilliet2005};
liquid fraction $\sim 5\; 10^{-2}$,
image width: 18 cm, courtesy C. Quilliet  (Univ. Grenoble).
}
\label{patterns_foam}
\end{figure}

\begin{figure}
(a)\includegraphics[width=6cm]{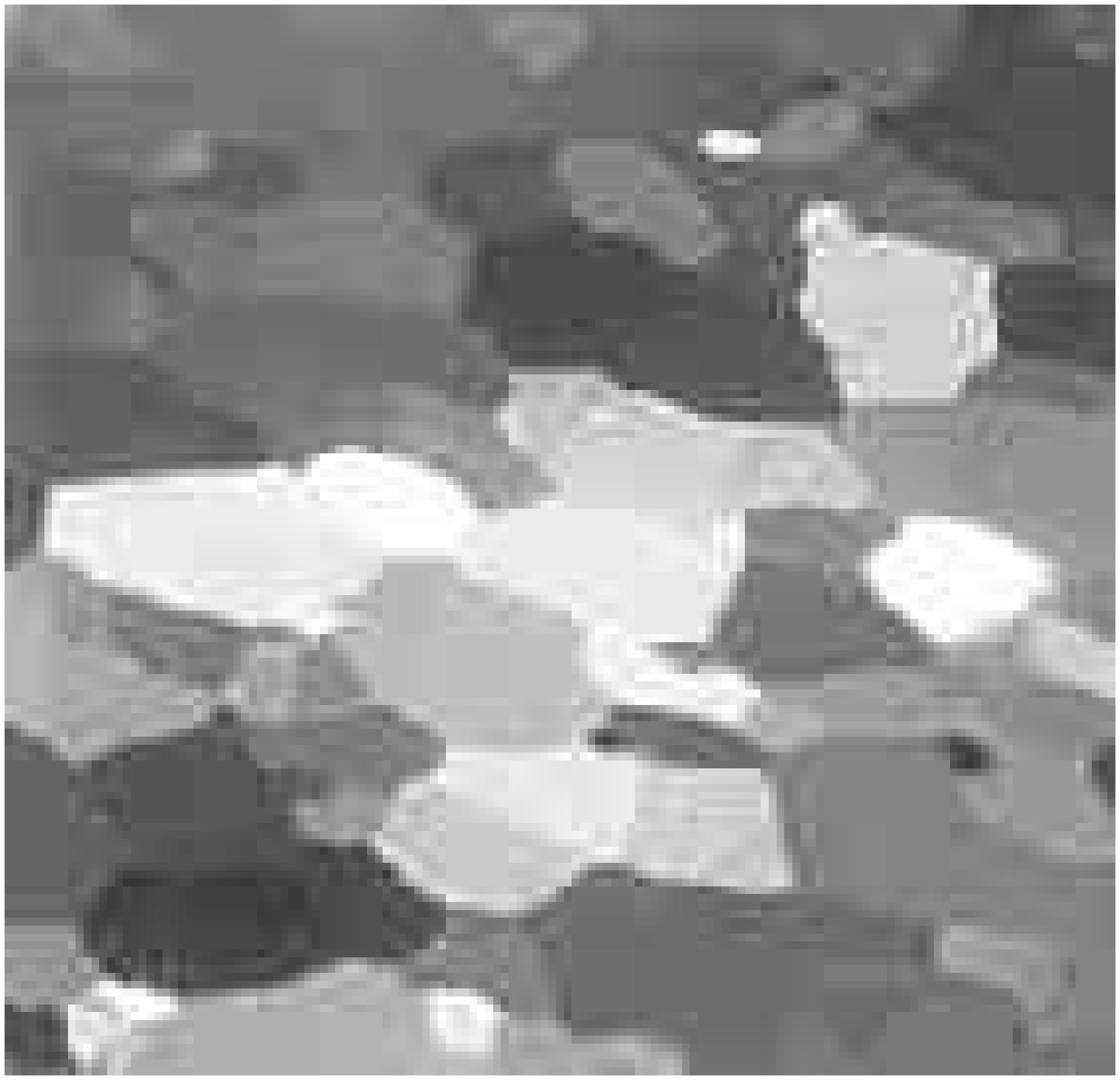}\\
(b) \includegraphics[width=6cm]{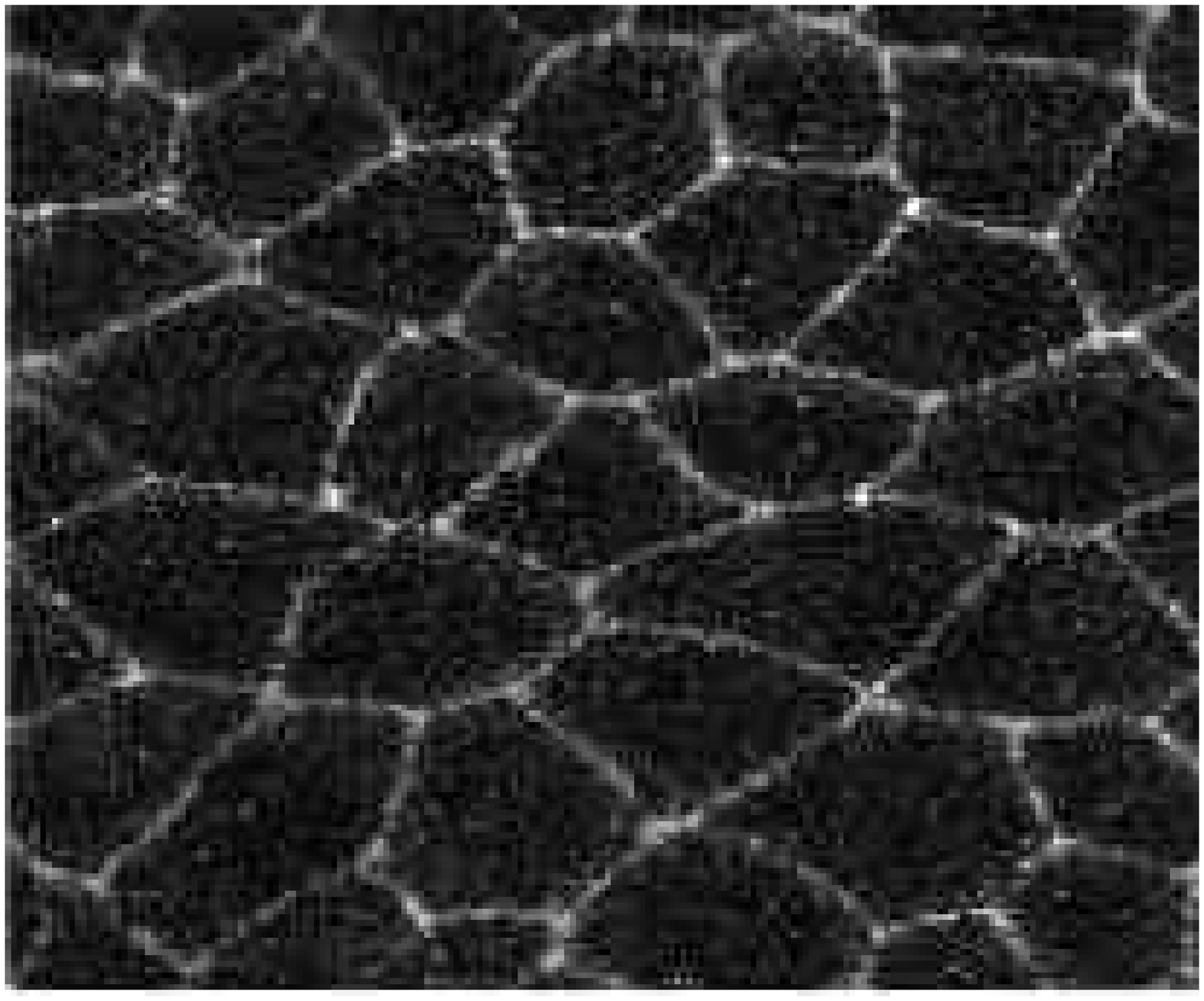}
\caption{
 Other cellular patterns. (a) Grains in a polycrystal of ice which rearranged during ice accumulation
\cite{durand2004}, image width: 10 cm, courtesy J. Weiss (Univ. Grenoble).
(b) Tissue of cells rearranging during the formation  of a fruit fly (Drosophila) embryo
 \cite{courtypreprint}: this thorax epithelium is labeled by the expression of the cell-cell adhesion
molecule E-Cadherin-GFP; image width: 160 $\mu$m,
 courtesy Y. Bella\"iche (Inst. Curie).
}
\label{patterns_cells}
\end{figure}

\begin{figure}
(a)
 \includegraphics[width=6cm]{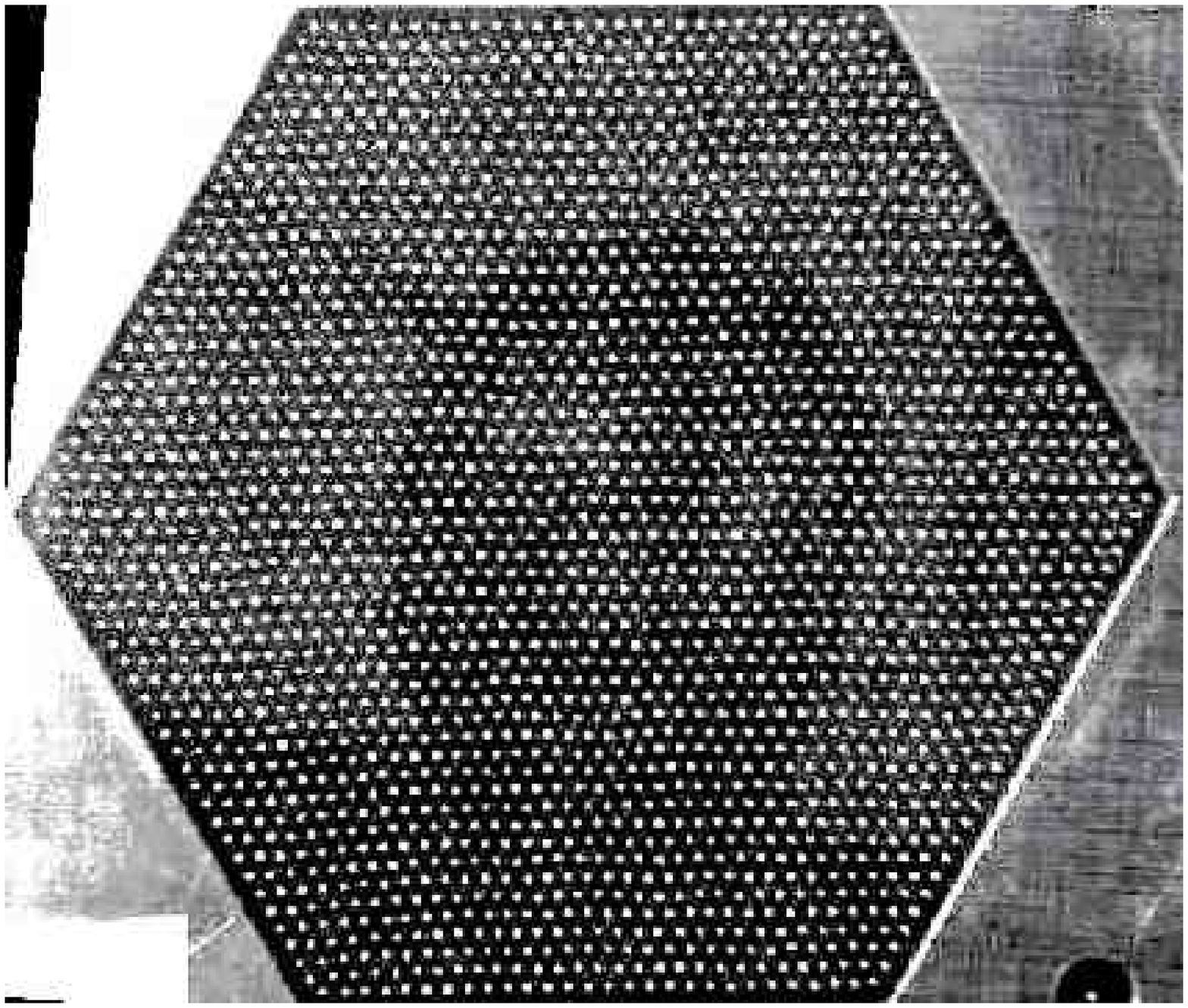}
 \\
(b)
\includegraphics[width=6cm]{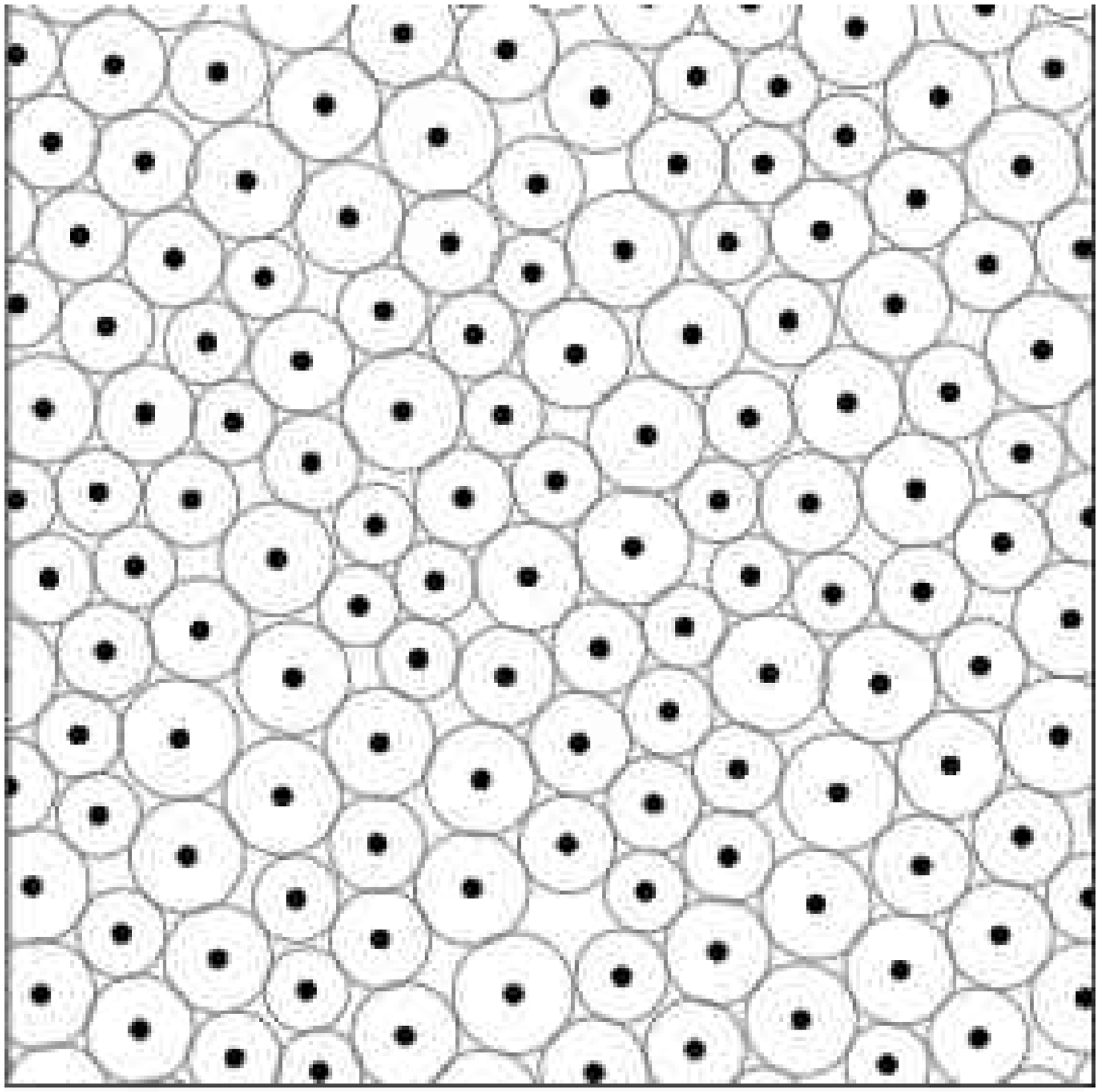}
\caption{
 Assemblies of particles. (a)
  Beads repelling each other; they are placed on a vibrating loudspeaker, with an effect shown to be equivalent to thermal fluctuations \cite{Cou06}; image size: 11.6 cm, courtesy G. Coupier (Univ. Grenoble)   \cite{Saint2004}.
  (b)
  Simulation of amorphous systems of atoms interacting via Lennard-Jones potential   \cite{Tanguy2006}: circles indicate each particle's effective radius (here with a 20\% dispersity), tangent circles correspond to vanishing interaction force;  image size: arbitrary, courtesy A. Tanguy (Univ. Lyon).
}
\label{patterns_particles}
\end{figure}


Cellular patterns include: liquid foams or emulsions (Fig. \ref{patterns_foam}); crystalline grains in polycrystals; or
biological tissues (Fig. \ref{patterns_cells}).
Assemblies of particles (Fig. \ref{patterns_particles}) include collections of beads,  molecules, or atoms; granular or colloidal materials; sets of tracers dispersed in a material, such as fluorescent probes or passively carried particles.
Despite their tremendous diversity of sizes and physical properties, all these patterns have a common point: they are made of a large number of well-identified individual objects.
We call them {\it discrete patterns}, where the word "discrete" here means the opposite of "continuous".
Other discrete patterns include interconnected networks, of {\it e.g.}  springs,  polymers, biological macromolecules, fibers, or telecommunication lines.

We define the pattern as {\it rearranging} if the mutual arrangement of the individual objects can change.  This is the case if they can move past each other, for instance due to
mechanical strain  (Figs. \ref{patterns_foam} or
 \ref{patterns_cells}a),
 spontaneous motility (Fig. \ref{patterns_cells}b),
or  thermal fluctuations
(Fig. \ref{patterns_particles}).
This is also the case if the number of individual objects can change, for instance due to  cell division or death
(Fig. \ref{patterns_cells}b), coalescence or nucleation of bubbles,  shrinkage during coarsening of polycrystals or foams.

\begin{table}
\begin{tabular}{|l|c|c|c|}
\hline
&&&\\
Pattern  & Texture  & Topological  & Geometrical \\
statistics & $\tensor{M}$  & changes & changes   \\
&
eqs. (\ref{eq:defM_lisible3D})
&
 $\tensor{T}$
&
 $\tensor{B}$
\\
&&eq. (\ref{eq:defT}) & eq. (\ref{eq:defB})\\
&&&\\
\hline
&&&\\
Statistical & Statistical & Statistical & Statistical\\
relative    & internal & topological & symmetrised \\
deformations & strain & rearrangement & velocity \\
&  $\tensor{U}$ & rate $\tensor{P}$ & gradient $\tensor{V}$ \\
&eq. (\ref{eq:defU})&eq. (\ref{eq:defP})& eqs. (\ref{eq:defG},\ref{eq:defV})\\
&&&\\
\hline
&&&\\
Continuous  & Current  & Plastic      & Total \\
medium & elastic  & strain  & strain  \\
strain & strain   & rate         & rate \\
& $\eps_{el}$ & $\deps_{pl}$ & $\deps_{tot}$ \\
&&&\\
\hline
\end{tabular}
\caption{Symmetric matrices used in the text. Equation numbers  correspond to their definitions. For comparison, the last row indicates the strains defined in continous mechanics for elastic, plastic and fluid behaviours.
}
\label{table}
\end{table}


Stimulated by the various imaging techniques, ref. \cite{mecke} reviews many tools available to describe and quantitatively characterise a pattern (that is, a single image). Another tool is
the  texture: it
appears in various contexts, including the order parameters of nematics,
the microstructure of polymers, or the fabric of grains; and has been used to describe mechanical strains by Aubouy {\it et al.}
 (see \cite{aub03} and references therein).
 It describes statistically how the individual objects are arranged with respect to each other.
With a few simple measurements performed directly from an image, it extracts quantitative information relevant to the size and anisotropy of the pattern. Since it is based on statistics, it is particularly useful for {\it disordered} patterns.
Here, we present a coherent set of robust tools (listed in Table \ref{table}), with a triple goal.

First, we revisit existing definitions of the texture   $\tensor{M}$, as well as the statistical strain  $\tensor{U}$  \cite{aub03} and the rearrangements  $\tensor{T}$  \cite{dollet_local} based on it. In fact, cellular patterns are better characterised using cell centers than using their  vertices. This remark enabled ref. \cite{dollet_local} to define a preliminary version of $\tensor{T}$. Here, we show that it also yields a more general definition of the texture, valid for all patterns. In addition, since a cell center is measured as an average over all pixels in a cell (while a vertex is a single pixel), measurements in experiments or simulations are more robust. The companion paper \cite{Cartes2D} shows that
displacements of cell centers (but not of cell vertices) are close to affine displacements. In order to make this paper self-contained, we recall and hopefully clarify the definitions of  $\tensor{M}$ and  $\tensor{U}$. We also present a more general definition of $\tensor{T}$ and derive explicitly its prefactor.

Second, thanks to their more general definition, we embed these three tools in a self-consistent formalism, which includes three additional ones: $\tensor{B}$, $\tensor{V}$ and $\tensor{P}$.
From two successive images in a movie, we extract information regarding the magnitude and direction of strain rate and rearrangements.

Third, we show that the six tools together provide a direct correspondence between a small scale, where $\tensor{M}$, $\tensor{B}$ and $\tensor{T}$ quantify the discrete pattern's local distortion and rearrangements; and a large scale, where $\tensor{U}$, $\tensor{V}$ and $\tensor{P}$ help describe a material as a continuous medium without any details
related with the discrete scale.
This enables to formulate elastic, plastic, fluid behaviours in a common, self-consistent modelling using continuous mechanics even for a discrete material. Experiments, simulations and theories can be expressed in the same language to be directly compared.


The only requirement is that the image should be of sufficient
quality to extract the positions of the centers of each individual
object (cell or  particle); as well as the list of neighbour pairs
(which objects are neighbours). All tools here are either static
or kinematic, and rely on the image only; that is, they are
independent of dynamics (stresses, masses and forces). They apply
to discrete patterns regardless of the size of their individual
objects, which can range from nanometers to meters or more. They
regard  simulations as well as experiments, and should enable
quantitative comparison between them. They apply  whatever the
pattern's disorder is.

Our equations are valid in any dimensions. For clarity, we write them in 3D, and show that is is straightforward to
rewrite them in 2D, see section \ref{2Dcase}.
We specifically choose to illustrate this paper with 2D images
(Figs. \ref{patterns_foam}-\ref{patterns_particles}), which are simpler and more common that 3D data.


More precisely, we illustrate each definition on the example of a foam flow (Fig. \ref{patterns_foam}a)  \cite{rau07}, which is both our original motivation and  the most suitable example.
Nitrogen is blown into water with commercial dishwashing liquid. Bubbles enter a channel, of length  1 m (only partly visible on the picture), width 10 cm, and thickness 3.5 mm: a monolayer of bubbles forms
 (area $A_{bubble}=16.0$ mm$^{2}$), sandwiched between two  glass plates
(quasi-2D foam, liquid fraction less than a percent). It steadily flows from left to right
 without vertical component (true 2D flow) until it reaches the free end of the channel.
Coalescence and ageing are below detection level.
A 3 cm diameter obstacle is inserted into the foam channel. The foam is forced to flow around it, resulting in a spatially heterogeneous velocity field. Different regions simultaneously display different
velocity gradients, internal strains, and rearrangement rates, and allow to
sample simultaneously many different conditions. Bubbles naturally act as tracers of all relevant quantities; and on the other hand the foam's overall behaviour appears continuous.
The total strain rate is partly used to deform bubbles and partly to make them move past each other; the companion paper \cite{Cartes2D} studies how it is shared between both contributions.


Sections \ref{sec:ingredients} and  \ref{sec:defM} start from  the static description of Ref.  \cite{aub03} and
develop it step by step, for pedagogical purpose, while refining it.
  Section \ref{sec:defBT} describes the changes between two successive images.
  Section \ref{sec:defUGP}  is useful to compare measurements on different patterns; or to compare experiments and simulations.
Section \ref{sec:micromacro} is more theoretical and regards specific applications: it discusses how to characterize  materials which behave as continuous media,
that is, where  the quantities  vary smoothly with space; and when it is possible to identify  our statistical  tools
with the usual quantities of continuous mechanics.
Appendices cover many practical,
 technical or theoretical details including all standard definitions
  and notations of matrices used in this paper.

\section{Texture and time evolution of links in the discrete pattern }
\label{sec:discrete_tensors}

\subsection{Ingredients}
\label{sec:ingredients}

The pattern is a collection of individual objects.
Here we are interested in the relative positions of these objects, not in each object's shape (although both are related in particular cases such as cellular patterns, see Appendix \ref{inertia}). We thus replace each object by a point called "site".

The user should adapt the measurement tools to the pattern under consideration, and the scientific questions
to be answered. For that purpose, the user should begin by deciding:
(i) what are the relevant links, that is,  pairs of sites which are connected (section \ref{sec:links});
 and (ii) the averaging procedure (section \ref{sec:moyennage} and Appendix \ref{average}).

These choices are conventions, and thus rather free. The results of the measurements depends on the chosen definition, but they are
much more robust than scalar measurements (see section
\ref{sec:defT}). Moreover, as
long as the same definition is used for all measurements, the equations that relate the
measurements  of the different quantities (such as eq. \ref{time_evolution_M}) are valid independently of the chosen definition. Once
conventions are chosen,  it is thus important to use them
consistently.

\subsubsection{Links between neighbouring sites}
\label{sec:links}

\begin{figure}
\includegraphics[width=9.5cm]{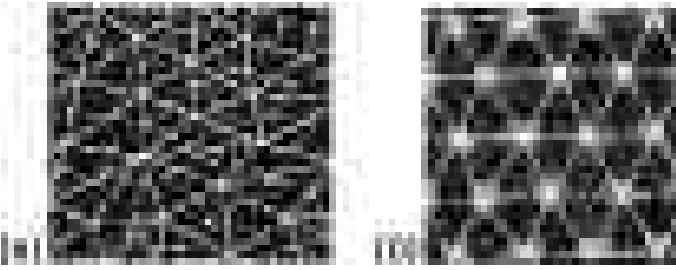}
\caption{Definition of sites and links.
(a) Cellular pattern. Background: detail from Fig. (\ref{patterns_cells}b). Foreground: a site is a cell's geometrical center; there is a link between two centers if their cells touch.
(b) Particle assembly. Background: detail from Fig. (\ref{patterns_particles}a). Foreground: a site is a particle's position; the links are defined as discussed in the text (here a Delaunay triangulation).
\label{sites}}
\end{figure}

In a {\it cellular pattern} (Figs. \ref{patterns_foam},\ref{patterns_cells}), it is often advisable to choose as sites each cell's geometrical center (see Fig. \ref{sites}a). However, alternative choices exist. For instance, a user interested in studies of dynamics might prefer the center of mass, if different from the geometrical center. Similarly, a biologist might be more interested in the cell's centrosome or nucleus.
Note that we do not advise to use a definition based on vertices (see Appendix
 \ref{vertex}).

When two
cells  touch each other it  defines that their sites are connected. This is unambiguous if cells walls are thin. This is the case for grains in polycristals, cells in an epithelium, or in a  foam with low amount of water (Figs. \ref{patterns_foam}a,\ref{patterns_cells}).

If cells walls are thick, as is the case in a foam with a higher amount of water (Fig. \ref{patterns_foam}b), different definitions of neighbours are possible. For instance, two cells are defined as neighbours if their distance is smaller than a given cut-off.
Or, if they are neighbours on a skeletonized image; that is, after an image analysis software has reduced cell walls to one pixel thick black lines on a white background.
 If cell walls are too thick,  cells are really separated (as in a bubbly liquid, where bubbles are round and far from each other) and can be treated like the particles, which we now discuss.

If each object is a {\it particle} (as in Fig.
\ref{patterns_particles}) it is natural to choose its center as
site (see Fig. \ref{sites}b). There are various possible choices
for the links. Since the tools characterize patterns and not forces, the definition of links is independent of interactions: a link between sites does not mean that sites interact; conversely, sites which interact are not necessarily linked. Whatever the chosen definition, it is important
that each particle has only a finite number of neighbours. 

In a first case (Fig. \ref{patterns_particles}b), the average distance between particles is comparable to their average radius; for instance, for a dense (also called compact or jammed) colloid or granular material.
We then recommend to define that two particles are linked if
their distance is less than a chosen cut-off. For hard spheres, this cut-off should be the sphere's radius plus a small tolerance.

In the opposite case,
  the average distance between particles is much larger than their average radius (Fig. \ref{patterns_particles}a); for instance, for a decompacted  colloid or granular material. We then recommend to recreate a cellular pattern by attributing to each particle its
 Voronoi domain (the set of points surrounding this particle, closer to it than to any other particle).
One then chooses to define that two particles are linked if their Voronoi domains touch; this is called the
"Delaunay triangulation" of the particles.

If the pattern is a {\it network}, it is natural to choose the nodes as sites. The connexions are physically materialised, and thus unambiguously defined.

\subsubsection{Averaging}
\label{sec:moyennage}

The present tools aim at describing the collective properties of links.
In what follows, $\left\langle .\right\rangle $ denotes the
average over a set of links relevant to the user:
 $$\left\langle .\right\rangle = N_{tot}^{-1} \sum (.),$$ where the sum is taken over
the number $N_{tot}$ of such links.
Appendix \ref{average} presents some technical details, especially regarding the boundaries of the averaging region, which can be treated as sharp or smooth.

The scale of study determines the number of links included.
Performing the same analysis at different scales (Fig. \ref{ElasticDeformationMap}) enables to obtain multi-scale results
\cite{durand2004,dollet_local,jan05}. For instance,  we can measure the dependence of pattern fluctuations with scale for particle assemblies.

Choosing to average over a small number $N_{tot}$  of links yields access to detailed local information.
For instance,
the local heterogeneity of a sample of ice  can be measured by
including the links around one single grain, then performing a comparison between different grains
\cite{durand2004}.
Similarly, to study the anisotropy of a biological cell which divides,
one can include only the links starting at this cell's center  \cite{courtypreprint}.

On the other hand, choosing a large number $N_{tot}$  of links yields better statistics.
This is the case for instance if the system is homogenous in space. In a homogenously sheared foam
 (Fig. \ref{patterns_foam}b), it makes sense to consider that all bubbles play a similar role, and average over the whole foam.
 Averaging over all links contained in the whole image enables to detect the overall anisotropy of an ice sample or an epithelium,  and compare it with other samples   \cite{durand2004,courtypreprint}  .

Even if the system is invariant along only one direction of space, one can average over this direction
\cite{jan05}.
Similarly, in a flow  which is invariant in time, one can
average over time (Fig. \ref{patterns_foam}a). In what follows,
 figures
are prepared with $1000$ successive
video images, representing  millions of bubbles: a time average yields good statistics and details of local variations, even if only a small part ({\it i.e.}  few links) of each image is included (Fig. \ref{ElasticDeformationMap}c).

\subsection{Texture $\tensor{M}$: current state of the pattern.}
\label{sec:defM}

We include this section, already published   \cite{aub03,dollet_local}, in order to make the present paper self-contained.

\subsubsection{Definition and measurement}
\label{defMbasic}

A pair of neighbour sites of coordinates $\vec{r}_1=(x_1,y_1,z_1)$ and $\vec{r}_2=(x_2,y_2,z_2)$ constitutes a link. Reducing the pattern to a set of links  sets aside the detailed information regarding the actual positions  of each site.

The link vector:
\begin{equation}
\ellperso=\vec{r}_2-\vec{r}_1,
\label{def_site_link}
\end{equation}
has coordinates  $(X,Y,Z) =  (x_2-x_1,y_2-y_1,z_2-z_1)$.
It carries the information on link  length and angle.
However,  $\ellperso$ and $-\ellperso$ play the same physical role: an average over several $\ellperso$s will yield a result which depends on this arbitrary choice of sign (and, in practice, if there are enough links, the average $ \left \langle \ellperso \right \rangle $ turns out to be close to zero).

 On the other hand, the number $\ell^2 = X^2 + Y^2 + Z^2$ is invariant under the change $\ellperso  \to -\ellperso$ and thus has a physically relevant (and non-zero) average:
$ \left \langle \ell^2 \right \rangle  =  \left \langle  X^2 + Y^2 + Z^2 \right \rangle$.
It reflects the average square link length, but loses the information of angle.

The link matrix $\tensor{m}$
  combines the advantages of both:
 \begin{equation}
 \tensor{m} =
 \left(
\begin{array}{ccc}
X^2&XY&XZ\\
YX&Y^2&YZ\\
ZX&ZY&Z^2
 \\  \end{array}
\right)
.
\label{eq:brut_lisible3D}
\end{equation}
 Its trace is ${\rm Tr} \left(  \tensor{m} \right)=
\ellperso^2$.
Its average defines the texture \cite{aub03}:
\begin{equation}
\tensor{M} =
  \left \langle \tensor{m}   \right \rangle
  =
\left(
\begin{array}{ccc}
 \left \langle X^2 \right \rangle & \left \langle XY \right \rangle & \left \langle XZ \right \rangle \\
 \left \langle YX \right \rangle & \left \langle Y^2 \right \rangle & \left \langle YZ \right \rangle \\
 \left \langle ZX \right \rangle & \left \langle ZY \right \rangle & \left \langle Z^2 \right \rangle
\\  \end{array}
\right)
.
\label{eq:defM_lisible3D}
\end{equation}
It is expressed in m$^2$. As required, it stores the same information regarding the current pattern: the square length, readily visible as the sum of diagonal terms; the angle and magnitude of anisotropy, as discussed below.

\subsubsection{Diagonalisation and representation}
\label{diago}

\begin{figure}
  \includegraphics[width=9cm]{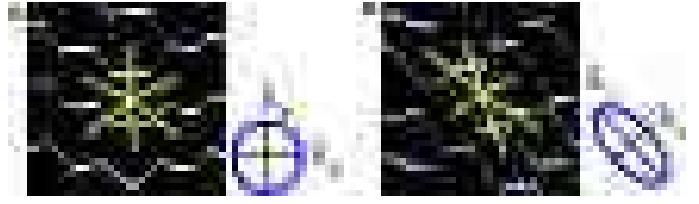}
\caption{Measurement of texture.
 Snapshots of two regions selected in Fig. (\ref{patterns_foam}a):
the foam is nearly isotropic in A, not in B.
 From the statistical analysis of
links (lines), and time average over several images, we compute the corresponding $\tensor{M}$. We
represent it  by an ellipse with axes proportional
 to the eigenvalues $\lambda_i$: in A it is nearly circular.
Thin lines indicate the axes with positive eigenvalues ({\it i.e.} here all axes).
}
\label{ElasticDeformation}
\end{figure}

By construction, $\tensor{M}$ is a matrix with symmetric off-diagonal terms ($XY=YX$, etc...).
It can thus be diagonalised (see Appendix \ref{diagonale} for details):
\begin{equation}
{\rm diag} \;  \tensor{M} =
\left(
\begin{array}{ccc}
\lambda_1 &0&0\\
0&\lambda_2&0\\
0&0&\lambda_3
\\  \end{array}
\right)
.
\label{eq:Mdiagonal}
\end{equation}
Its three eigenvalues $\lambda_i$ ($i=1$, 2 or 3) are
positive. Their sum,  Tr$\tensor{M}$,  is exactly  $ \left \langle \ell^2 \right \rangle$. In practice, they usually  have the same order of magnitude, of order of $ \left \langle \ell^2 \right \rangle /3$.

In a truly 3D pattern, $\tensor{M}$ has strictly positive eigenvalues
(except in unphysical examples). Thus its inverse $ \tensor{M}^{-1}$ always exists (eq. \ref{eq:Sinverse}).


$\tensor{M}$ can be represented as an ellipsoid, which axes directions are that in which $\tensor{M}$ is diagonal, represented as thin lines on Fig. (\ref{ElasticDeformation}).
Each ellipsoid's axes length is proportional to the corresponding $\lambda_i$. It is expressed in m$^2$:
this is less intuitive than m, 
\label{meter} 
and ellipsoids are more elongated that the actual cell shape; but this is necessary for the consistency with the representation of the other matrices (see Appendix \ref{inertia}). 
The square link length  $ \left \langle \ell^2 \right \rangle$ is reflected in the size of the ellipsoid, more precisely as the square root of the sum of the three axes lengths; it is thus {\it not} proportional to the ellipsoid's volume.
The direction in which links are longer is represented by the direction of ellipsoid elongation: the greater the pattern's anisotropy, the more elongated the ellipsoid.
If the texture is measured at several regions of the image, it is represented as several ellipsoids, that is, a map of the texture field $ \tensor{M} (\vec{R},t) $
(see also section \ref{2Dcase} and Fig. \ref{ElasticDeformationMap}).

When the pattern is statistically isotropic, so is its texture. It is thus diagonal in any system of axes, and the three $\lambda_i$s are exactly equal,
$ \lambda_i = \left \langle \ell^2 \right \rangle /3$:
\begin{eqnarray}
 \tensor{M} &\stackrel{\rm isotropic}{=}&
\left(
\begin{array}{ccc}
\frac{\left \langle \ell^2 \right \rangle}{3} &0&0\\
0&\frac{\left \langle \ell^2 \right \rangle}{3}&0\\
0&0&\frac{\left \langle \ell^2 \right \rangle}{3}
\\  \end{array}
\right)
\nonumber \\
&=&
\frac{\left \langle \ell^2 \right \rangle}{3}\; \left(
\begin{array}{ccc}
1 &0&0\\
0&1&0\\
0&0&1
\\  \end{array}
\right)
.
\label{eq:Misotrope}
\end{eqnarray}
That is, the texture of an isotropic pattern contains only the information of length: $\tensor{M} = \left \langle \ell^2 \right \rangle \tensor{I}_3 / 3$, where $ \tensor{I}_3$ is the identity matrix in 3D. It is represented as a sphere. In that case, all axes are equivalent (or "degenerated").

\subsubsection{Two-dimensional case}
 \label{2Dcase}

If the pattern under consideration is contained in a plane, as are most experimental images, we turn to a 2D notation.
As mentioned, this is straightforward:
\begin{equation}
2D: \quad \quad \tensor{M} =
\left(
\begin{array}{cc}
 \left \langle X^2 \right \rangle & \left \langle XY \right \rangle\\
 \left \langle YX \right \rangle & \left \langle Y^2 \right \rangle
\\  \end{array}
\right)
.
\label{eq:defM_lisible2D}
\end{equation}
  There exist two orthogonal axes (eigenvectors) in which  $\tensor{M}$ would be diagonal:
\begin{equation}
2D: \quad \quad {\rm diag} \;  \tensor{M} =
\left(
\begin{array}{cc}
\lambda_1 &0\\
0&\lambda_2\\
  \end{array}
\right)
.
\label{eq:Mdiagonal2D}
\end{equation}
with  strictly positive $ \lambda_i $, ($i=1$ or 2).
 If we call $  \left \langle \ell_+ \right \rangle$ the r.m.s. length of links in the direction of elongation (say, 1) and
$  \left \langle \ell_- \right \rangle$ the r.m.s. length of links in the direction of compression (say, 2), then
$ \lambda_1 \approx \left \langle \ell^2_+ \right \rangle /2$ and $ \lambda_2 \approx \left \langle \ell^2_- \right \rangle /2$.
 Its inverse $ \tensor{M}^{-1}$ always exists.

\begin{figure}[hbtp]
(a) \includegraphics[width=7cm]{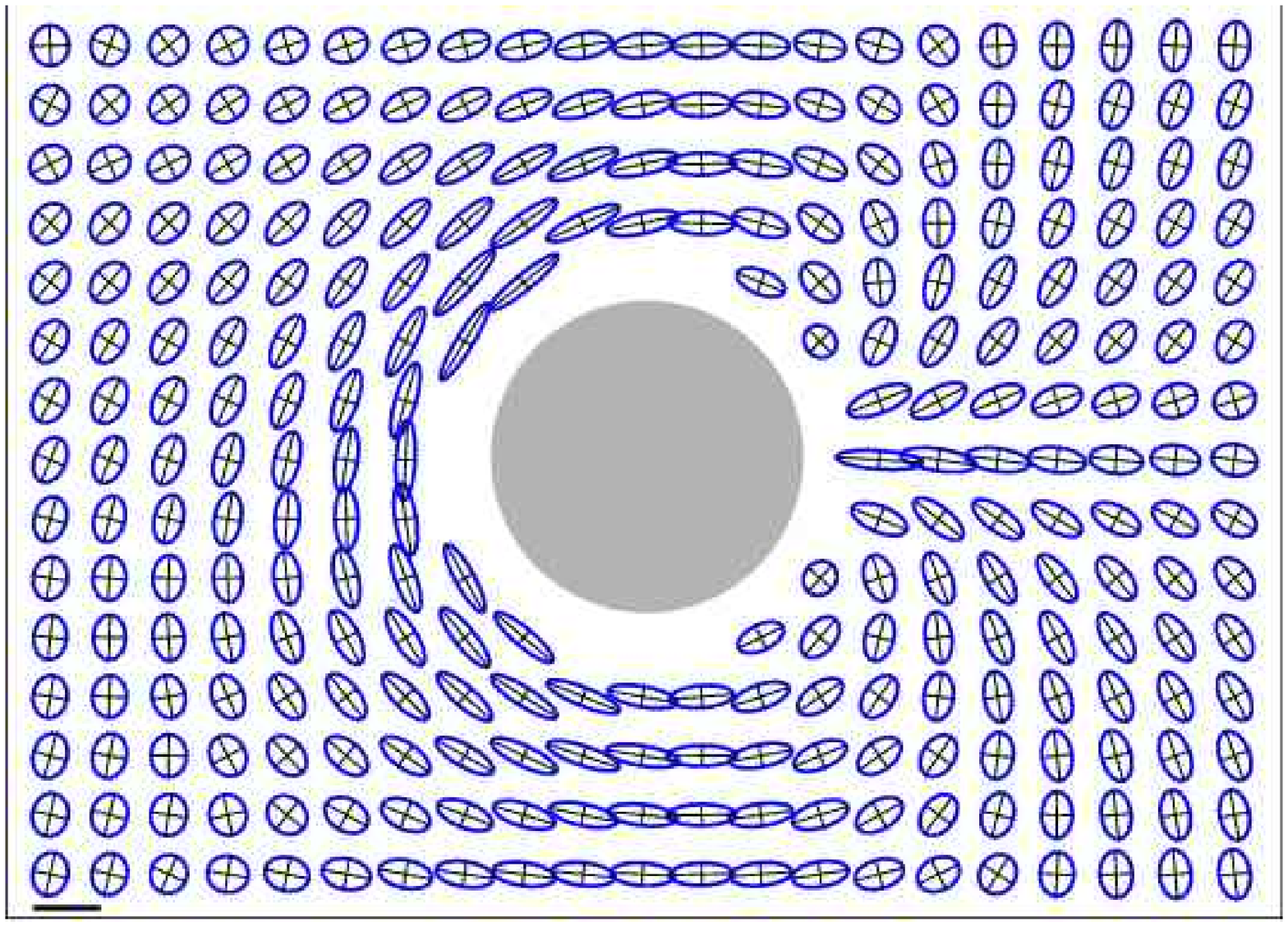}\\
  (b) \includegraphics[width=7cm]{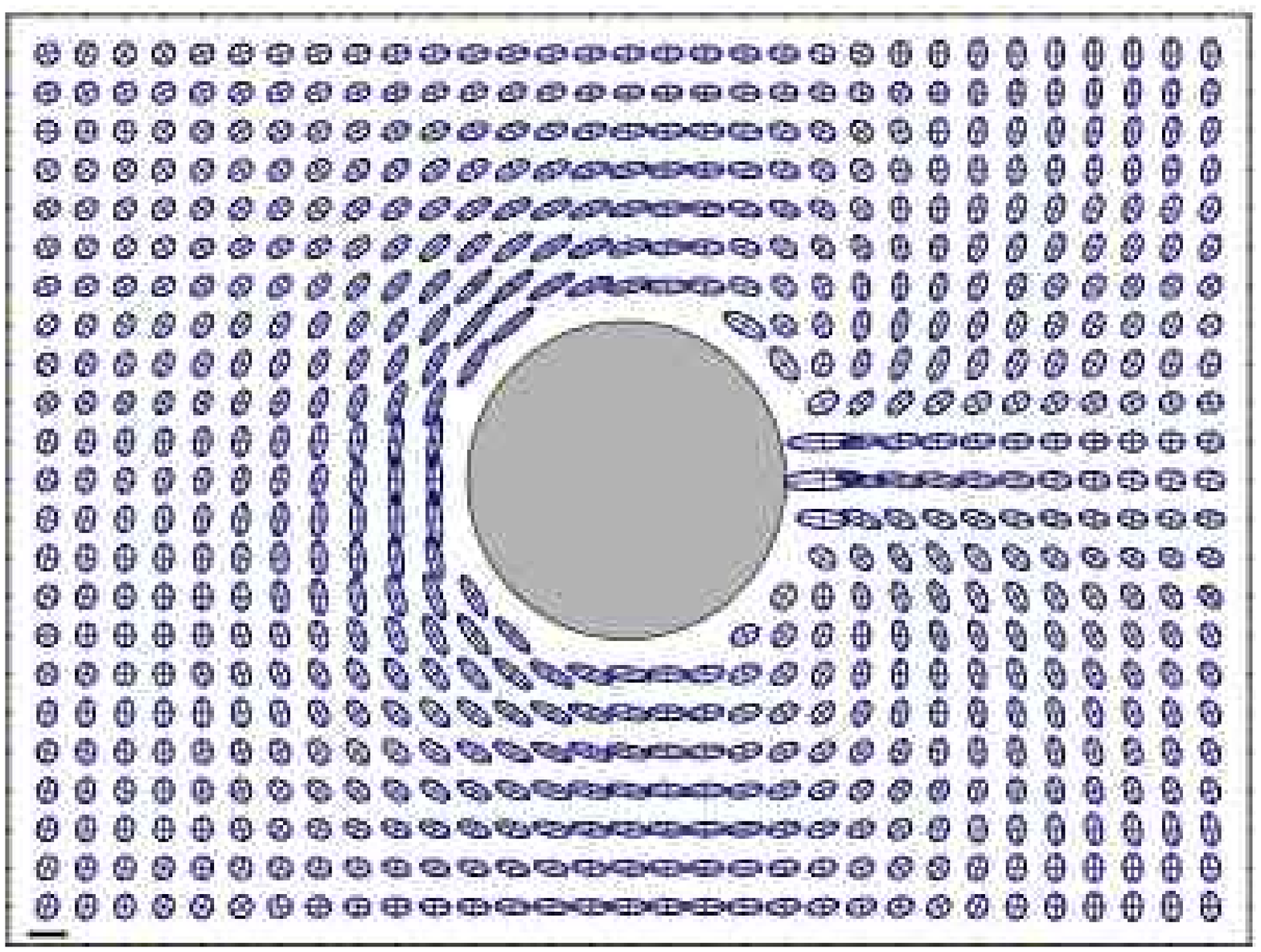}\\
  (c)\includegraphics[width=7cm]{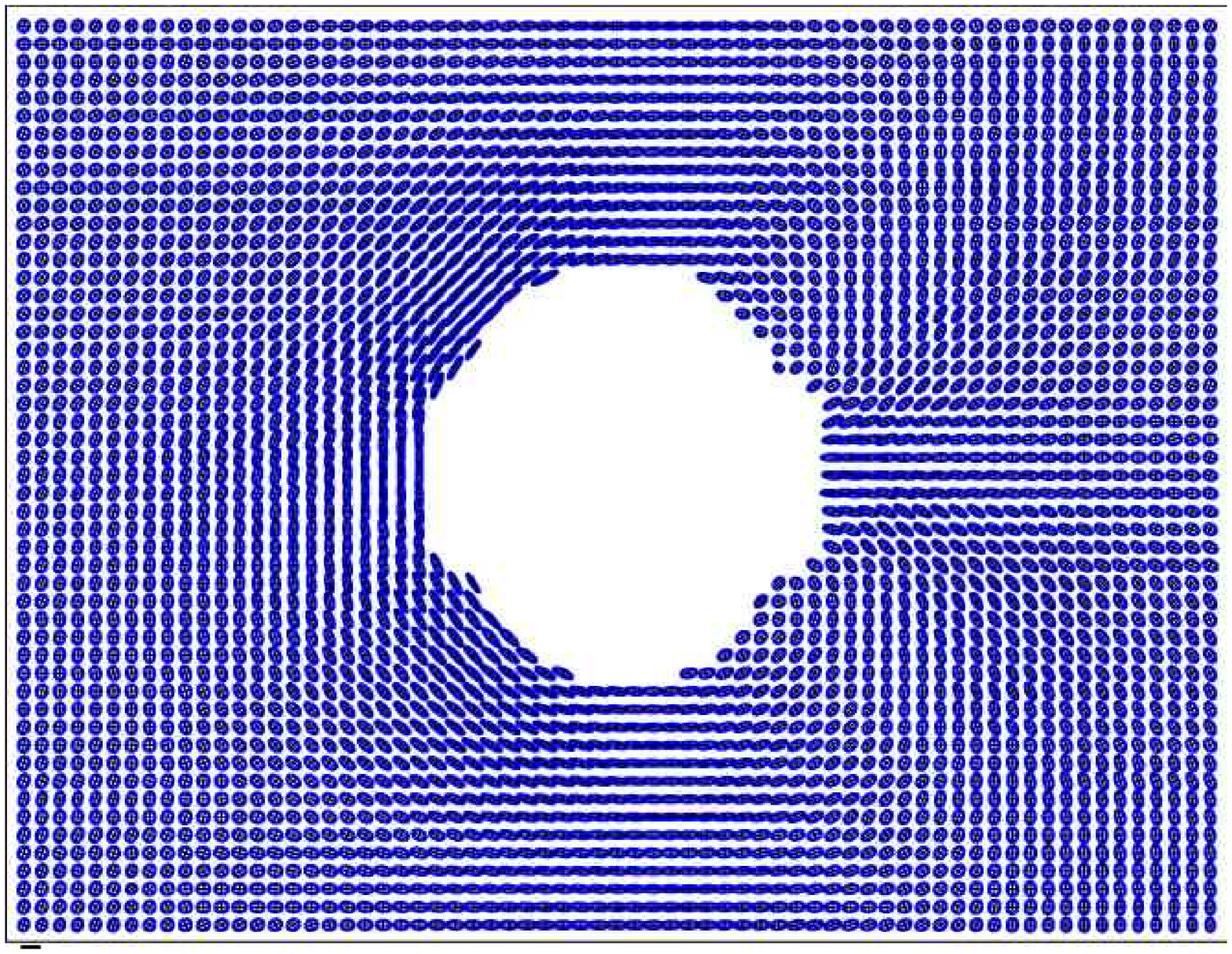}\\
 \caption{Map of the texture measured in each region of the foam flow (Fig. \ref{patterns_foam}a).  The area $V_{box}$ of each averaging box corresponds to (a) 3, (b) 1 and (c) 0.3 bubbles (sub-bubble resolution); since there are 1000 movie images, and 3 times more links than bubbles \cite{wea99}, this corresponds to averages over $10^4$, $3\; 10^3$ and $10^3$  links, respectively. Scale:
actual image size, 15 cm $\times$ 10 cm; $\tensor{M}$ (ellipses): bar $=$ 10 mm$^2$, for the eigenvalues, which are here all positive, and represented as ellipse axes lengths.}
\label{ElasticDeformationMap}
\end{figure}

In 2D, $\tensor{M}$ is represented by an ellipse  (Fig. \ref{ElasticDeformation}).
  Measurements can be performed at larger scale to decrease the noise due to fluctuations, or at smaller scale to evidence more details of the spatial variations (Fig. \ref{ElasticDeformationMap}). 
  
When the pattern is isotropic, its texture is diagonal with any choice of axes:
 \begin{eqnarray}
  \tensor{M} & \stackrel{\rm 2D}{\stackrel{\rm isotropic}{=}}&
\left(
\begin{array}{cc}
\frac{\left \langle \ell^2 \right \rangle}{2}&0\\
0&\frac{\left \langle \ell^2 \right \rangle}{2}\\
 \end{array}
\right)
\nonumber \\
&=&
\frac{\left \langle \ell^2 \right \rangle}{2}\; \left(
\begin{array}{cc}
1 &0\\
0&1\\
 \end{array}
\right)
,
\label{eq:Misotrope2D}
\end{eqnarray}
That is, $  \tensor{M} = \left \langle \ell^2 \right \rangle  \tensor{I}_2/2$,
where $ \tensor{I}_2$ is the identity matrix in 2D.
All axes are equivalent (or "degenerated") and the angle of eigenvectors is not defined.
$  \tensor{M} $ is represented by a circle and the thin lines lose their signification
(Fig. \ref{ElasticDeformation}A).

\subsection{Time evolution}
\label{sec:defBT}


Differentiating eq. (\ref{eq:defM_lisible3D}) determines how $\tensor{M}$ varies.
Appendix \ref{time_evolution}, useful for practical calculations, discusses finite size effects due to the  time interval between successive images of a movie. Neglecting these effects in eq. (\ref{divided})
yields the simplified time evolution of $\tensor{M}$:
\begin{equation}
\frac{\partial\tensor{M}}{\partial t}
+ \tensor{M} \frac{\partial \log N_{tot}}{\partial t}
= -\vec{\nabla} \cdot {\cal J}_{M}+\tensor{B}+\tensor{T}.
\label{time_evolution_M}
\end{equation}
The variation of $N_{tot}$ is negligible in most physical examples; however, it is significant for instance in biological tissues with many divisions \cite{courtypreprint} or in coarsening systems such as ageing foams \cite{wea99,weairerivier}.

 The three terms of the r.h.s.  can be measured on
a movie, and have the following meaning. In the time interval  between two successive images, some links enter or exit the region of averaging; some links change their length or angle; some links are created or destroyed, respectively.
They
are now discussed one by one (sections \ref{sec:defJ}, \ref{sec:defB} and \ref{sec:defT}; respectively).

\subsubsection{Flux ${\cal J}_{M}$ }
\label{sec:defJ}

In eq. (\ref{time_evolution_M}),  ${\cal J}_{M}$ is the flux of advection, that is, the  transport of texture.
It counts the rate at which links enter or exit throught the sides of the region of averaging.
Technically, it is a rank-three tensor ({\it i.e.} with 3 indices): for more details see ref. \cite{dol05} and  Appendix \ref{outerprodu}.
In a good approximation, ${\cal J}_{M}\simeq  \vec{v} \otimes \tensor{M}$,
where $\vec{v}$ is the local average
velocity, see Appendix \ref{time_evolution}.
 Its divergence
$\vec{\nabla} \cdot {\cal J}_{M}$
counts the net balance between links that enter and exit; it
  vanishes if
$\tensor{M}$ is spatially
homogeneous, or at least is constant along a flux
line; it also vanishes if the local average
velocity is zero.

\subsubsection{Geometrical texture changes: $\tensor{B}$}
\label{sec:defB}

\begin{figure}[hbtp]
\includegraphics[width=5cm]{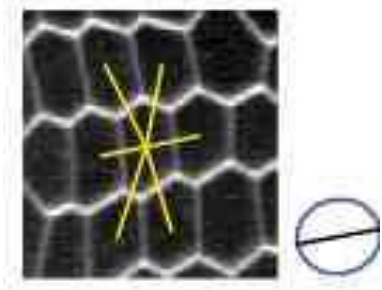}
\caption{
Changes in the shape of links (geometry).
In a first snapshot of a small region of the foam, links are represented by dashed lines.
The same bubbles are tracked on a next image, with links represented by solid lines.
To evidence how each link changes, we have removed the overall translation (which plays no role here), and superimposed both snapshots. We calculate $ \tensor{B}$ and plot it as a coffee bean: an ellipse with a thin line indicating the positive eigenvalue (direction of extension).
}
\label{fig:ContinuousGrowth}
\end{figure}

 $\tensor{B}$ describes the changes in the pattern's overall shape, that is, geometry:  at which rate, and in which direction, the pattern deforms.
 It   reflects relative movements: it is insensitive to a global, collective translation.

 $\tensor{B}$ is based on the  links which,  on both successive images,  exist ({\it i.e.} do not undergo topological rerrangement) and belong to the region of averaging  ({\it i.e.} are not advected).
These links may change in length and direction (Fig. \ref{fig:ContinuousGrowth}).
We obtain each link's
 contribution directly from
eq. (\ref{eq:brut_lisible3D}) and average it, like in eq. (\ref{eq:defM_lisible3D}):
\begin{equation}
\tensor{B} =
   \left\langle
        \frac{{d}\tensor{m} }
             {{d}t}
    \right\rangle
    .
\label{eq:defB}
\end{equation}
Appendix \ref{time_evolution} provides more details; it also defines  $\tensor{C}$, which symmetrical part is related with $\tensor{B}$, and will turn useful to define $\tensor{W}$ and $\tensor{V}$ in section \ref{sec:defG}.
The right hand side of eq. (\ref{eq:defB}) should not be confused with
${d} \left\langle \tensor{m} \right\rangle / {d}t  $; that is,  $\tensor{B}$  is not equal to the variation of  $\tensor{M}$: as eq. (\ref{time_evolution_M}) shows, the difference between them is due to changes in the links included in the averages.

 $\tensor{B}$  is symmetric. Its units are in m$^{2}$s$^{-1}$.
 It
has a positive eigenvalue in a direction of extension, and a negative one in a direction of compression.  In case of dilation, all its eigenvalues are positive.
In a region of shear, it is plotted as an ellipse with one thin line drawn on it, similar to a "coffee bean"
(Fig. \ref{fig:ContinuousGrowth}). We can plot a map of  $\tensor{B}$: it is similar to  Fig.
(\ref{fig:tauxdef}), data not shown.

\subsubsection{Topological texture changes: $\tensor{T}$}
\label{sec:defT}

\begin{figure}[hbtp]
(a)\setlength{\unitlength}{1cm} \begin{picture}(8,2.7)(0.0,0.0)
\put(0,0){\includegraphics[width=8.5cm]{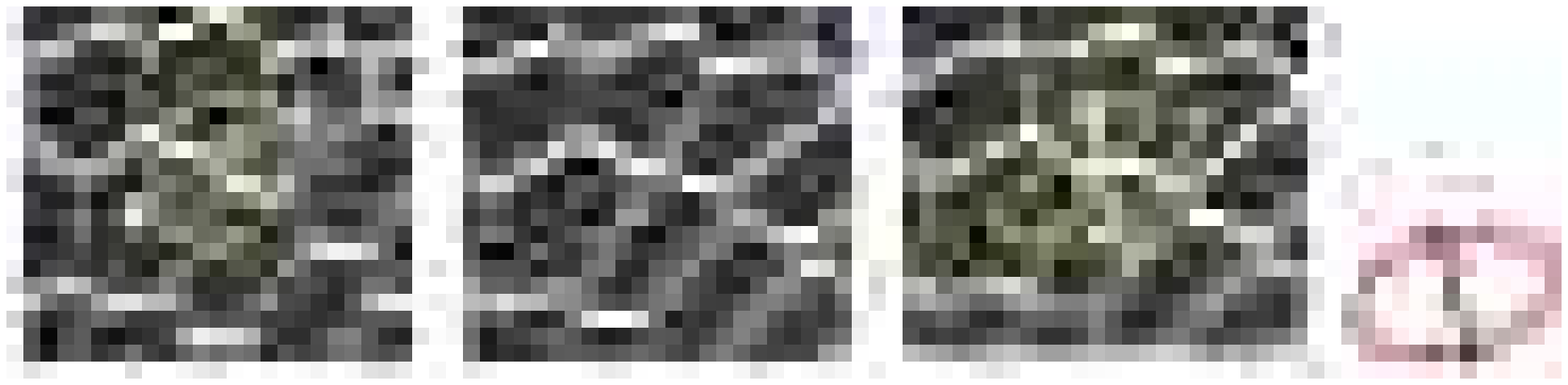}}
\put(-0.5,1){$\ellperso_{d}$} \put(6.5,2.5){$\ellperso_{a}$}
\end{picture}
\\[3mm]
 (b)\includegraphics[scale=0.7]{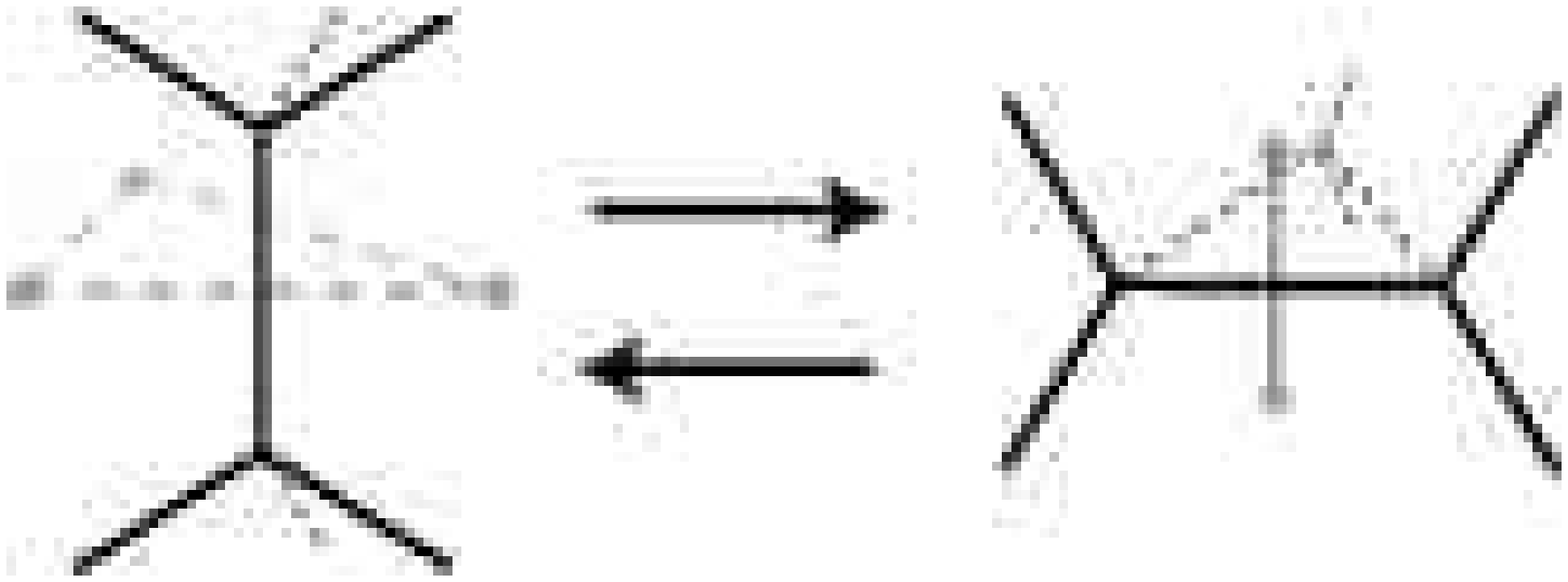} \\[3mm]
(c)\includegraphics[scale=0.7]{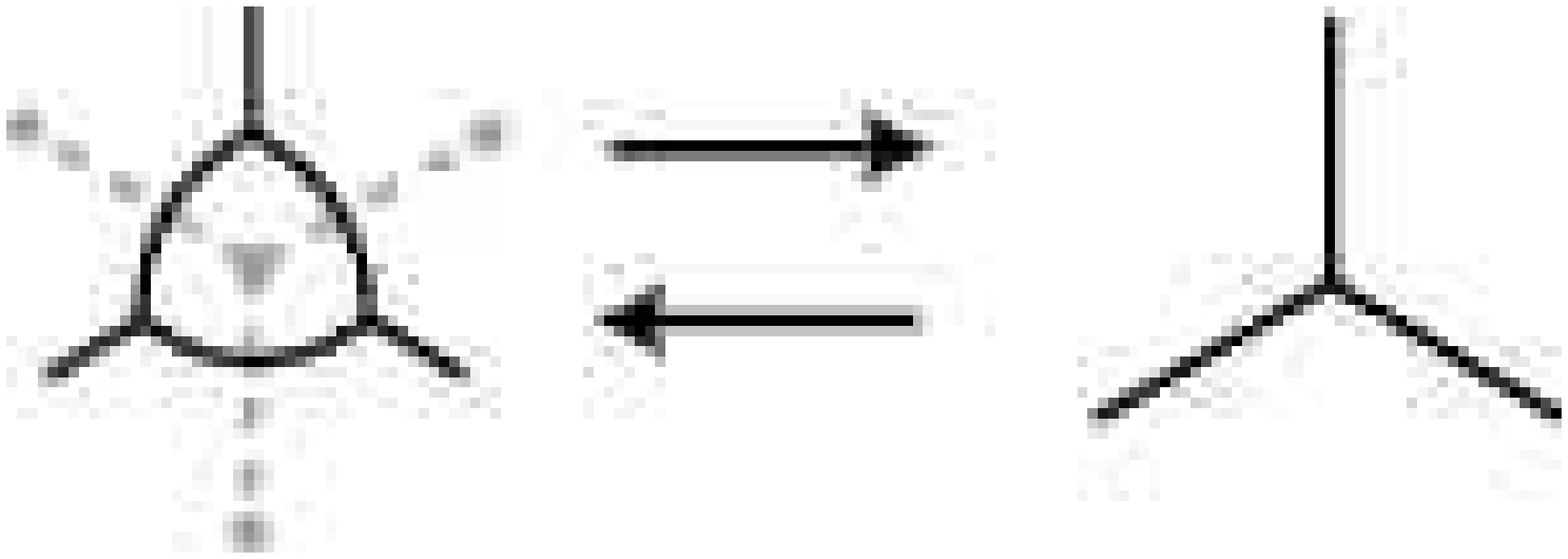} \\[3mm]
(d)\includegraphics[scale=0.7]{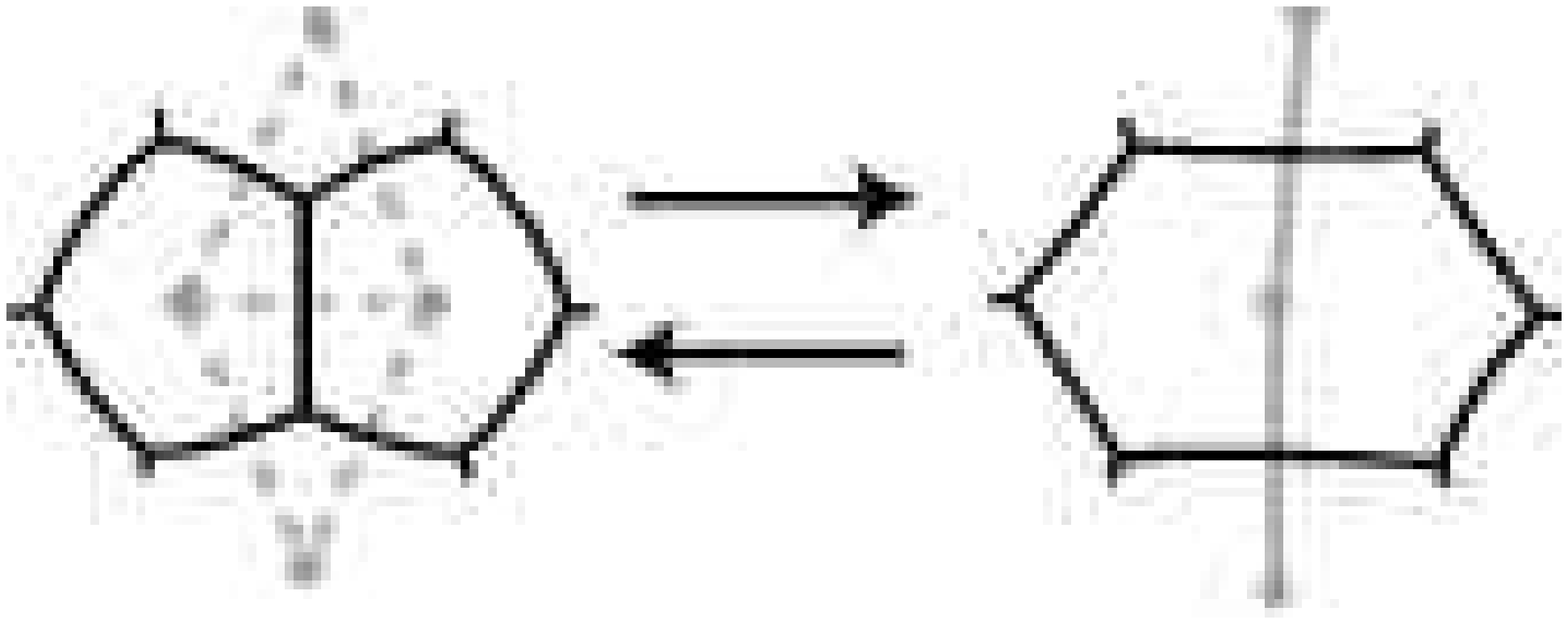} \\
\caption{
Changes in the list of links (topology). They are here illustrated by cellular patterns but apply to all other patterns as well.
(a) Neighbour exchange in 2D: snapshots extracted from a dry foam \cite{Cartes2D}, one link disappears (dashes) and another appears (thick grey line); and corresponding representation of $\tensor{T}$ as an ellipse, where a thin line indicates the positive eigenvalue. (b) Same,
sketched  in 3D: three links (hence 3 faces) disappear and one appears. (c) Site disparition, in 2D or 3D: all its links disappear. (d) Coalescence of two sites, in 2D or 3D: the link between them disappears, the links to their common neighbours merge (here there is a total of five disappearances and two creations).}
\label{fig:Topologicalchanges}
\end{figure}

$\tensor{T}$ reflects the topological changes, namely, changes in the list of links:
creation and destruction, that is, source term of the texture.
Appendix \ref{time_evolution} provides details.
Briefly, each link $ \ellperso_{a}$ which has appeared since the preceding image
has a contribution
given by eq. (\ref{eq:brut_lisible3D}), noted $\tensor{m}_{a}$; and similarly
the contribution of
a link which disappears before the next image is noted  $\tensor{m}_{d}$.
Averaging over all links which appear or disappear between successive images
defines  $\tensor{T}$ as:
 \begin{equation}
\tensor{T}=\dot{n}_{a}\left\langle \tensor{m}_{a} \right\rangle -\dot{n}_{d}\left\langle \tensor{m}_{d} \right\rangle .
\label{eq:defT}
\end{equation}
The quantity $\dot{n}_{a}$  (resp. $\dot{n}_{d}$),  expressed in s$^{-1}$,
 is {\it not} the time derivative of a physical quantity (which we would note $d/dt$).
It is the rate of link appearance (resp.
disappearance), per unit time
and per existing link.
If $\dot{n}_{a}$   and $\dot{n}_{d}$ are equal, their inverse is
 the average link's life expectancy.

$\tensor{T}$ is expressed in m$^{2}$s$^{-1}$. It characterises the total effect on the pattern of all topological changes occurring between two images.  By construction it is symmetric, like $\tensor{M}$: it can thus be diagonalized and represented as an ellipsoid.
It is robust to artefacts and errors in determinations of neighbours \cite{dollet_local}.
It is general, and includes
information of frequency, size, direction and anisotropy for all contributions of all topological changes: they can be treated indifferently and added together.
However, as we now discuss, the user might be interested in studying separately the contributions of the different processes
\cite{courtypreprint}.

The coalescence of two sites (Fig. \ref{fig:Topologicalchanges}d) corresponds in foams to the breakage of a liquid wall between two bubbles, with a net balance of minus one site \cite{ohl98}. The reverse process corresponds in epithelia to a cell division, and results in one more site \cite{dub98}.
When the number of sites decreases (resp. increases), so does the number of links, and $\tensor{T}$ usually has only negative (resp. positive) eigenvalues. The variation in the number of sites and links is thus visible in the trace of $\tensor{T}$.

Neighbour exchanges (Fig. \ref{fig:Topologicalchanges}a,b), also called "T1" in  the case of cellular patterns
\cite{wea99,weairerivier},
preserve the sites;
in 2D, they also preserve the number of links, $\dot{n}_{a}=\dot{n}_{d}$.
It is for that case that ref. \cite{dollet_local} introduced a specific definition of $\tensor{T}$.
$\tensor{T}$ usually is mostly deviatoric (Appendix \ref{not_tens}), with both positive and negative eigenvalues (and a vanishing trace), in directions correlated with the appearing and disappearing links. Note that the eigenvectors are exactly orthogonal, while the appearing and disappearing links need not be:
thus eigenvectors are not strictly parallel to links, especially when $\tensor{T}$
is measured as an average over several individual topological processes.
For our flowing foam example, a map of $\tensor{T}$  (data not shown) would be similar to  Fig. 18 of ref.  \cite{dollet_local}, or to Fig.
\ref{fig:Pmeasured} below.

 If a pattern has a free surface, when sites exchange neighbours the number of links can vary, and thus locally $\tensor{T}$ can be far from deviatoric.
\label{CK} 

The disparition of a site (Fig. \ref{fig:Topologicalchanges}c) corresponds in foams to a bubble
which shrinks, also called "T2"
\cite{wea99,weairerivier}; and in epithelia to a cell which dies, or exits the epithelium plane. The reverse process is a site nucleation. Both processes have an approximately isotropic contribution to $\tensor{T}$.

\section{Statistical tools to obtain relative deformations and their time evolution}
\label{sec:defUGP}

This section facilitates comparison between different experiments; or between
experiments, simulations and theory. This is useful for large scale deformations and flows: 
e. g. of particle
assemblies, of foams and emulsions, of granular materials, or of  biological
tissues during morphogenesis.

Here we try to link the discrete, local description of  Section \ref{sec:discrete_tensors} with the
continuous, global description of  Section \ref{sec:micromacro}.
For this continuous description to
be self-consistent, it is necessary to
  get rid  of the discrete objects' length scale,
 that is, the typical size of links.
For each of the three
discrete quantities $\tensor{M}$, $\tensor{B}$, $\tensor{T}$
defined in section \ref{sec:discrete_tensors}, it is possible to
  construct a continuous counterpart: that is, a tool
which is dimensionless (or expressed in s$^{-1}$),
with no m$^2$ any longer. This defines $\tensor{U}$, $\tensor{V}$, $\tensor{P}$, respectively
(sections \ref{sec:defU},  \ref{sec:defG} and  \ref{sec:defP}, respectively).

\subsection{Statistical internal strain : $\tensor{U}$}
\label{sec:defU}

The internal strain has been defined by Aubouy {\it et al.} \cite{aub03}
through a comparison between the current pattern and a reference one.
We include it here  in order to make the present paper self-contained, and to provide additional explanations and examples.

\subsubsection{Strain of a single link}

Consider first a link $\ellperso$ of length $\ell$, and apply to it an infinitesimal variation  $d\ell$.
Its relative extension, or infinitesimal strain, is $d\ell/\ell$,  or equivalently $d(\log\ell)$  \cite{tan03}.
The "true strain" (also called "Hencky strain" \cite{tan03}) is defined with respect to a state $\ellperso_0$
chosen as a reference (often a state without stress) using several equivalent expressions:
\begin{eqnarray}
\int \frac{d\ell}{\ell} &= &\log\left( \frac{\ell}{\ell_0}\right)
=
\frac{1}{2}  \log\left( \frac{\ellperso^2}{\ellperso_0^2}\right)
\nonumber \\
&=&
 \frac{1}{2} \left[  \log {\rm Tr} \left(  \tensor{m} \right) -   \log {\rm Tr} \left(  \tensor{m}_0 \right)  \right].
\label{strain_link}
\end{eqnarray}
We perform these manipulations because the last expression of eq. (\ref{strain_link}) is the easiest to generalise. It is not a problem to take the log of  dimensioned quantities (here, the square of a length) because this cancels out in the final result.

\subsubsection{Statistical strain of the pattern}

For a whole pattern, replacing $ {\rm Tr} \left(  \tensor{m} \right)$ by  $\tensor{M}$
enables to perform statistical averages over links.
The logarithm of $\tensor{M}$ is unambiguously defined and is easily performed in three standard steps on a computer (Appendix \ref{diagonale}). It suffices to first, switch to  the  three orthogonal axes ($\tensor{M}$'s eigenvectors) in which $\tensor{M}$ is diagonal; second, take
the logarithm of its  eigenvalues, which are  strictly positive  (section \ref{diago}):
\begin{equation}
{\rm diag} \; \log \left(  \tensor{M} \right) =
\left(
\begin{array}{ccc}
\log \lambda_1 &0&0\\
0& \log \lambda_2&0\\
0&0&\log \lambda_3
\\  \end{array}
\right)
;
\label{deflogM}
\end{equation}
and third, switch back to the original axes.
It is necessary to perform first all linear operations such as averaging.
This ensures in particular that all $\lambda_i$s in eq. (\ref{deflogM}) are non-zero.
Taking the logarithm, which is a non-linear operation, has to be performed later.

Eq. (\ref{deflogM}), like  eq. (\ref{strain_link}), requires to define a reference,
 expressed in the
same units as $\tensor{M}$, so that the difference of their logarithms is
well defined and dimensionless.
Such a reference  texture $\tensor{M}_0$ is discussed in section \ref{M0_text}.

The "statistical internal strain" is defined  \cite{aub03}  as:
\begin{equation}
  \tensor{U}
  =  \frac{1}{2} \left(  \log \tensor{M} - \log   \tensor{M}_0 \right) .
\label{eq:defU}
\end{equation}
Here $ \tensor{U}$  completely characterises the material's current strain:
relative dilation, amplitude and direction of anisotropy.

\subsubsection{Reference texture  $\tensor{M}_0$ }
\label{M0_text}

  Practical details regarding the reference texture  $\tensor{M}_0$ are presented in
appendix  \ref{M0_appendice}.

Eq. (\ref{eq:defU}) shows that the exact choice of $\tensor{M}_0 $ affects the value
 of $ \tensor{U}$  but not its variations. It thus does not appear explicitly in the kinematics
 (eqs. \ref{eq:Vshared},\ref{kine}) nor in the dynamics (for instance in the value of the shear modulus,
 appendix  \ref{shear_modulus}).
Moreover,   eq. (\ref{eq:defU})  remains unchanged if we multiply both $\tensor{M}_0 $ and $\tensor{M}$
by a prefactor; this is why the exact unit ({\it e.g.} m$^2$, mm$^2$, $\mu$m$^2$) in which  $\tensor{M}_0 $ and $\tensor{M}$ are expressed is unimportant, as long as it is the same unit for both.\label{prefactor}

Whatever the choice, the reference is defined by the texture $\tensor{M}_0 $.
It  suffices to determine 6 numbers (3 numbers, if in 2D); or 1, in the (most common) case where $\tensor{M}_0 $ is isotropic.
Only the reference texture corresponding to the current state plays a role; past changes of the reference pattern, for instance during an irreversible strain (also called "work hardening" \cite{cha87}), need not be taken into account.
It is never necessary to know the details of the corresponding pattern's structure, such as the positions of each object one by one: it is even not necessary that this pattern exists and is realisable.

\subsubsection{Examples}
\label{ExU}

\begin{figure}[t]
\includegraphics[width=8cm]{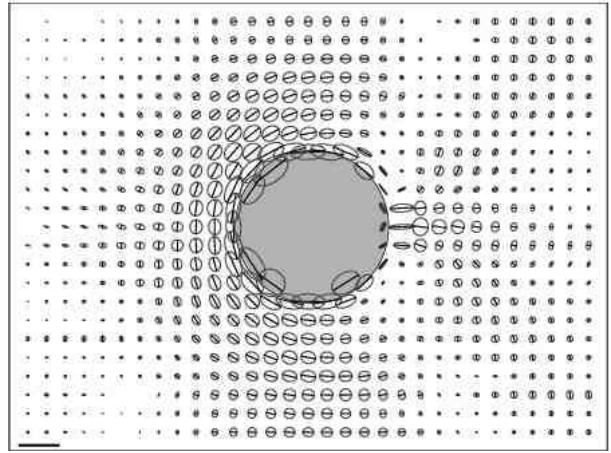}
\caption{ Elastic behaviour. Map of  the statistical internal strain $\tensorstrain$   (eq. \ref{eq:defU}) measured on the foam of Fig. (\ref{patterns_foam}a). Coffee bean
axes indicate the direction and amplitude of stretching (indicated
by a thin line) and compression. $ \tensor{M}_0$ is chosen as the averaged value of $ \tensor{M}$ measured at the left and right of the image, far from the obstacle.  Same box size as in Fig. (\ref{ElasticDeformationMap}b).
Scale: for ellipses  axes lengths, bar = $1$ (dimensionless) for the positive eigenvalue and the absolute value of the negative one.
\label{cap:elastic} }
\end{figure}

This section presents a few examples and particular cases of internal strain.

(i) If the material is uniformly dilated (affine deformation, see section \ref{affine}) by a factor $k$ in all directions, then $\tensor{M}=k^2\tensor{M}_0$.
Thus $\tensor{U} = \log(k) \tensor{I}_D$, as is expected for instance for gases; that is, in 3D:
$$\tensor{U} =
\left(
\begin{array}{ccc}
\log k &0&0\\
0& \log k &0\\
0&0& \log k
\\  \end{array}
\right)
.$$

(ii)
Conversely,  if the
material is uniformly dilated by a factor $k$  in one
direction and compressed by a factor $1/k$ in
another direction, then:
$$
{\rm diag} \; \tensor{U}  =
\left(
\begin{array}{ccc}
\log k &0&0\\
0& -\log k&0\\
0&0&0
\\  \end{array}
\right)
.
$$

(iii)
For incompressible materials,
$\tensor{U}$'s diagonal terms
are usually both positive and negative, and their sum is usually small:
$\tensor{U}$ is mostly deviatoric (Appendix \ref{not_tens}).
   Note that even in incompressible materials
    the links' mean square length can vary slightly,
so that Tr($\tensor{U}$) is not necessarily strictly zero.
For instance, it reaches  0.03 in  Fig. (\ref{patterns_foam}b) where a foam is sheared while keeping bubble number and total foam area exactly constant (Ataei Talebi and Quilliet, private communication).

(iv) In Fig. (\ref{cap:elastic}),
most ellipses look circular; deviations from circles occur close to the obstacle.
 We distinguish regions where extension dominates, and  ellipses  are stretched
like coffee beans, from regions where compression dominates, where the ellipses are flattened like capsules.

(v) If (but only if) $ \tensor{M}_0$ is isotropic, then
 $\tensor{U}$ is diagonal in the same axes
 as $\tensor{M}$.
Then eq. (\ref{deflogM}) enables to rewrite eq. (\ref{eq:defU}) more explicitly:
\begin{equation}
{\rm diag} \;   \tensor{U} =
\left(
\begin{array}{ccc}
\log \sqrt{\frac{\lambda_1}{\lambda_0}}  &0&0\\
0& \log  \sqrt{\frac{\lambda_2}{\lambda_0}} &0\\
0&0&\log  \sqrt{\frac{\lambda_3}{\lambda_0}}
\\  \end{array}
\right)
,
\label{defUcomplet}
\end{equation}
where $\lambda_0$ is $\tensor{M}_0$'s eigenvalue
(e.g.  $\lambda_0 = \left \langle \ell_0^2 \right \rangle/3$ is $\tensor{M}_0$ if we use the definition of eq. \ref{eq:Misotrope_pourM0}).
 Eq. (\ref{defUcomplet})
reflects that  $\tensor{M}$ and  $\tensor{U}$
have the  same
eigenvectors: they
commute. Eq. (\ref{defUcomplet})
 also  relates the trace of $\tensor{U}$ with $\tensor{M}$'s determinant (product of eigenvalues):
 \begin{eqnarray}
{\rm Tr} \;   \tensor{U} &=& \log \sqrt{\frac{\lambda_1\lambda_2\lambda_3}{\lambda_0^3}}
\nonumber \\
&=& \frac{1}{2}\log\;  \left( {\rm det} \;   \tensor{M} \right)
- \frac{1}{2}\log\;  \left( {\rm det} \;   \tensor{M}_0 \right)
.
 \end{eqnarray}

(vi) in the limit
of small strains, {\it i.e.} when $\tensor{M}$ remains close enough to  $\tensor{M}_0$,
eq. (\ref{eq:defU}) can be linearised   \cite{aub03}.
The difference of logarithms simply amounts to a division by
$\tensor{M_{0}}$, that is:
$\tensorstrain\simeq(\tensor{M}-\tensor{M}_{0})\tensor{M}_{0}^{-1}/2.$
   This is true whether  $\tensor{M}_0$ is isotropic or not
(unlike  eq. \ref{defUcomplet}). This approximation is used in
Appendix  \ref{upperconvectedderivative}.

\subsection{Kinematics: time evolution}

\subsubsection{Statistical velocity gradient: $\tensor{W}$ and $\tensor{V}$ }
\label{sec:defG}

We want to define the continuous counterpart  of the geometrical changes
$\tensor{B}$
(eq. \ref{eq:defB}). We use  $\tensor{M}^{-1}$ (eq. \ref{eq:Sinverse}), which is in m$^{-2}$, and is always defined.
For reasons which appear below (eqs.
\ref{link_under_affine}-\ref{G_affine}),  we  use $\tensor{C}$  (eq. \ref{eq:defC}) as an intermediate step,
and
 define
 $\tensor{W}$ as:
 \begin{equation}
    \tensor{W}
=
  \tensor{M}^{-1}
  \;
    \tensor{C}
=
\left\langle\ellperso\otimes\ellperso\right\rangle^{-1}
\;
\left\langle
\ellperso
\otimes
\frac{d\ellperso}{dt}
\right\rangle
    .
\label{eq:defG}
 \end{equation}
  $\tensor{W}$ has the
dimension of a strain  rate (s$^{-1}$): its order of magnitude
is the links' average variation rate.
Like $   \tensor{B} $, it vanishes when the pattern moves as a whole, with a rigid body translation.

For reasons which appear below (eqs.
\ref{link_under_affine}-\ref{G_affine}),
we  call it the "statistical velocity
gradient", and
we purposedly define it
 as
 $ \tensor{M}^{-1} \;   \tensor{C}$
 rather than
 $  \tensor{C} \; \tensor{M}^{-1}$.
In general,  $\tensor{{G}}$ is not symmetric.

\begin{figure}
{\includegraphics[width=8.5cm]{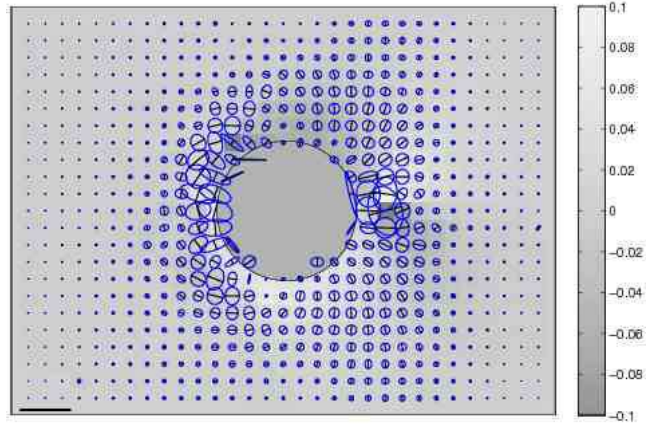}}
\caption{
Fluid behaviour.
  Map of the statistical symmetrised velocity
gradient $\tensor{V}$ (eq. \ref{eq:defV})
 measured on the foam of Fig. (\ref{patterns_foam}a). Coffee bean
axes indicate the direction and amplitude of stretching rate (indicated
by a thin line) and compression rate.
Same box size as in Fig. \ref{ElasticDeformationMap}b).
Scale: for ellipses  axes lengths, bar = $0.1$ s$^{-1}$  for the positive eigenvalue
         and the absolute value of the negative one.
 Grey levels: statistical  vorticity from the rotation rate $\tensor{\Omega}$ (eq. \ref{eq:defOmega}) in s$^{-1}$.
}
\label{fig:tauxdef}
\end{figure}

In practice, the most useful
quantity is
its symmetric part, the "statistical symmetrised velocity
gradient": \begin{equation}
\tensor{V}=\frac{\tensor{W}+\tensor{W}^{t}}{2}=
\frac{
\tensor{M}^{-1}\tensor{C}
+
\tensor{C}^{t}\tensor{M}^{-1}
}{2}.
\label{eq:defV} \end{equation}
It is the rate of  variation of
$\tensor{U}$ due to the links'
stretching and relaxation.
In cases where $\tensor{C}$ and $\tensor{M}$ commute, $\tensor{M}^{-1}\tensor{C}$ is symmetric and
eq. (\ref{eq:defV}) simply writes $\tensor{V}= \tensor{B}\tensor{M}^{-1}/2.$

Fig. (\ref{fig:tauxdef}) plots an example of $\tensor{V}$. It is large all around the obstacle,
but only very close to it; it is almost the same before and after
the obstacle. When the material's density is constant,  Tr$\tensor{V}$ is small (but not necessarily exactly zero), and the corresponding
ellipse is nearly (but not necessarily exactly) circular.

The anti-symmetric part is the statistical rotation rate: \begin{equation}
\tensor{\Omega}=\frac{\tensor{W}-\tensor{W}^{t}}{2}=
\frac{
\tensor{M}^{-1}\tensor{C}
-
\tensor{C}^{t}\tensor{M}^{-1}
}{2}.
\label{eq:defOmega}
\end{equation}
It has 3 independent components in 3D, but only 1 in 2D (appendix \ref{not_tens}).
Thus Fig. (\ref{fig:tauxdef}) plots it as grey levels.

\subsubsection{Statistical topological rearrangement rate: $\tensor{P}$}
\label{sec:defP}

\begin{figure}
 \includegraphics[width=7.4cm]{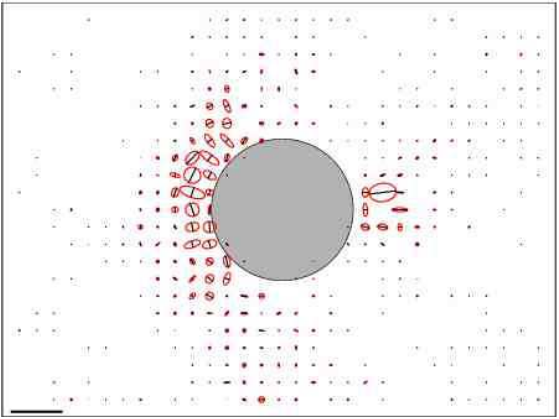}
\caption{ Plastic behaviour.
   Map of the topological strain rate $\tensor{P}$ (eq. \ref{eq:defP})
 measured on the foam of Fig. (\ref{patterns_foam}a). Coffee bean
axes indicate the direction of links which have just disappeared (indicated
by a thin line) and just appeared; note that this is the inverse of
Fig. (\ref{fig:Topologicalchanges}), due to the minus sign in eq. (\ref{eq:defP}).
Same box size as in Fig. \ref{ElasticDeformationMap}a). Measurement boxes touching the obstacle were removed. Scale: for ellipses  axes lengths, bar = $0.1$ s$^{-1}$  for the positive eigenvalue
 and the absolute value of the negative one, both  proportional to the frequency
of rearrangements.
}
\label{fig:Pmeasured}
\end{figure}

We  define
 the continuous counterpart  of the topological changes $\tensor{T}$
(eq. \ref{eq:defT}) in a way similar to
eq. (\ref{eq:defV}):
\begin{equation}
\tensor{P}\; =\; -\frac{1}{2}\; 
\frac{
\tensor{M}^{-1}\tensor{T}
+
\tensor{T}\tensor{M}^{-1}
}{2}
.
\label{eq:defP}
\end{equation}
Here we have introduced a factor $-1/2$ so that  $\tensor{P}$ is the term
which unloads the statistical internal strain, as will appear  in
eqs. (\ref{identify_plast},\ref{kine}).
In cases where $\tensor{T}$ and $\tensor{M}$ commute, such as in the companion
paper \cite{Cartes2D},  $\tensor{M}^{-1}\tensor{T}$ is symmetric and
eq. (\ref{eq:defP}) simply writes $\tensor{P}=-\tensor{T}\tensor{M}^{-1}/2.$

This "statistical topological rearrangement rate" $\tensor{P}$ (eq. \ref{eq:defP})
 has the
dimension of s$^{-1}$. It measures the frequency  and
direction of rearrangements: it
 is of the order of magnitude of the number of changes per unit time and per link.
Corresponding ellipses are elongated like coffee beans  (resp: flattened like capsules) if the number of links decreases (resp: increases); if the number of links is conserved, ellipses are nearly circular.

As an example, Fig. (\ref{fig:Pmeasured}) shows that the rearrangements are more
frequent just in front of the obstacle, or in a very narrow region
behind it. The rate of rearrangements decreases smoothly with the
distance to the obstacle. This is due to the foam's elasticity. It
contrasts with the sharp transition between solid-like and fluid-like
regions observed in purely visco-plastic materials \cite{ber85,bla97}.
The companion paper \cite{Cartes2D} presents an example with a larger spatial distribution of topological events, which enables for a better spatial resolution.

\subsubsection{Kinematic equation of evolution}

We have thus three independent symmetric matrices:
$\tensor{U}$,
$\tensor{V}$ and
$\tensor{P}$.
As discussed in Appendix \ref{kine_eq_obj}, there is a relation between them.

In the case where we can neglect the variation in $N_{tot}$ and the higher order terms in $\tensor{U}$, the time evolution of $\tensor{M}$
 approximately simplifies as (eqs. \ref{time_evolution_M},\ref{kine},\ref{eq:corotationnal}):
\begin{equation}
\tensor{V}=\frac{{\cal D}\tensor{U}}{{\cal D}t}+\tensor{P}.
\label{eq:Vshared} 
 \label{identify_plast} 
\end{equation}
That is,
 the (statistical) symmetrised velocity gradient is shared between two
contributions: one part (which includes advection and rotation) changes the (statistical) internal strain,
the other part is the (statistical) topological rearrangement rate.
How $\tensor{V}$  is shared between both contributions
constitutes the main subject
of the companion paper \cite{Cartes2D}.

Physically, eq. (\ref{eq:Vshared}) means that, when a perturbation is applied to the
overall shape of the pattern,
part of it affects the appearance of the pattern (loading) and the other part
goes into rearrangements (unloading).
Section  \ref{sec:micromacro} introduces a parallel point of view, in particular with
eq. (\ref{eq:total-strain-rate-is-shared}).

\section{Continuous mechanics}
 \label{sec:micromacro}

This section applies to materials which dynamics can be described using continuous mechanics
 (section \ref{continuous}), in terms of stresses.
We examine whether it is possible to relate  the continuous, large-scale, dynamical description on one hand; and on the other hand the  statistical measurements based on discrete objects introduced in section  \ref{sec:defUGP}, which describe the pattern's connections (topology), shape (geometry)  and movements (kinematics).

 We recall that continous mechanics involves three kinematical quantities
$\deps_{tot}$, $\eps_{el}$ and $\deps_{pl}$
 (section \ref{EVP}),
which are related through eq. (\ref{eq:total-strain-rate-is-shared}).
We then try (section \ref{link_discrete_continuous}) to   identify it with eq.  (\ref{eq:Vshared}).

\subsection{Continuous description and RVE}
\label{continuous}


If the material acts as a continuous medium
\cite{cha87,bat00,lan86}, it usually has the following  properties.
First, there exists a range of
$V_{box}$ sizes  over
which measurements
yield the same results  \cite{dollet_local}.
In that case, the box is called
a representative volume
element (RVE).
\label{RVE}
This is usually obeyed if $V_{box}$ is much larger than the range of interaction between individual objects, and also larger than the  correlation length of their disorder (but these conditions are neither necessary nor sufficient).
Second, its description can be local in space, that is, its equation of evolution involves partial space derivatives, and the spatial variations of its solutions look smooth.
Third, the average quantities have at large scale a role more important than that of fluctuations.


Regarding the choice of the RVE,
the discussion of section \ref{sec:moyennage} applies.
Here again, averages  $\left\langle .\right\rangle $ on detailed geometrical quantities
are performed on
a spatial box of volume $V_{box}$ and over a time $\tau$ selected to suit  the problem under
consideration.
The shape of the box  should preferably respect
the system's symmetries.

For the present purpose of a continuous description,
there is however the additional requirement  that $N_{tot}\gg1$. More precisely,
the relative statistical uncertainty
 $N_{tot}^{-1/2}$
should be smaller than the relative precision
required by the user.
  A few tens or hundreds of links are often enough
(there is no need for $10^{23}$ links).
This does not set any theoretical lower limit to the size of $V_{box}$ : it can well be as small as  the link size, or even smaller, if there are enough images to average (Fig. \ref{ElasticDeformationMap}c).

\subsection{Elastic, plastic, fluid behaviours}
\label{EVP}

If the pattern behaves as a continuous material,  we can consider a  RVE (section
 \ref{RVE})  at position $\vec{R}$. The   velocity field   is $     \left\langle \vec{v}\right\rangle (\vec{R})$, that is, an average over the whole RVE.
If
$\vec{R}_1$ and $\vec{R}_2$ are the positions of
two RVEs,  the velocity gradient
$ \tensor{\nabla v}$ is  the spatial derivative of the velocity field, and $ \tensor{\nabla v}^t$
is its transposed (eqs. \ref{defgradv},\ref{defgradv_transposed}), then:
\begin{equation}
      \left\langle \vec{v}\right\rangle \left( \vec{R}_2 \right)
   \simeq
    \left\langle \vec{v}\right\rangle \left( \vec{R}_1\right)
    +
 \tensor{\nabla v}^t
  \cdot
   \left( \vec{R}_2 - \vec{R}_1\right)
   .
\label{velograd}
\end{equation}
Details on this notation can be found in Appendix \ref{outerprodu}.
Eq.   (\ref{velograd})   neglects  terms
of order of $  \left| \vec{R}_2 - \vec{R}_1\right|^2 $
and higher.
It describes the velocity field as continuous and  affine, that is,  a
term which varies linearly with position plus a constant term (offset).

One of the key ingredients of continuous mechanics is
the velocity gradient's symmetrical part, that is, the   total strain rate:
\begin{equation}
\deps_{tot}=
\frac{
 \tensor{\nabla v}+\tensor{\nabla v}^{t}
 }{2}
.
\label{deftotaldefo}
\end{equation}
This is a purely kinematical quantity, but it determines the
contribution to the viscous (dissipative) stress  \cite{bat00}.

For small strain (linear elastic regime), neglecting advection and
rotation, the integration of eq. (\ref{deftotaldefo}) defines a total applied
strain,
which is a function of the past history of the sample, as:
 \begin{equation}
\tensor{\eps}_{tot}
=
\int dt\;
\deps_{tot}
\approx
\frac{
 \tensor{\nabla u}+\tensor{\nabla u}^{t}
 }{2}
.
\label{Landauluimeme}
\end{equation}
Here $\tensor{\nabla u}$ is the gradient of the
 displacement field $\vec{u}$, and $\tensor{\deps}_{tot}$
 its symmetrical part.

The total strain rate $\tensor{\deps}_{tot}$ contributes in part (loading) to change the elastic strain $\eps_{el}$, and in part (unloading) to a plastic strain rate  $\deps_{pl} $ which is defined by their difference:
\begin{equation}
\deps_{tot}=\frac{{\cal D}\eps_{el}}{{\cal D}t}+
\deps_{pl}.
\label{eq:total-strain-rate-is-shared}
\end{equation}

Alternatively elasticity and plasticity 
 are defined through dynamics. A given region of the pattern is said to be in elastic, plastic or viscous  regime, according to the
contribution to the stress that dominates locally
\cite{pinceau}. 
The elastic strain
$\eps_{el}$ contributes to the reversible part
of the stress. Plasticity
describes the irreversible contribution to the stress in the low velocity
limit (note that rearranging patterns can often deform a lot without breaking).
Both are solid behaviours, that is, exist in the limit of very low velocity gradient.
The viscous contribution to stress is irreversible: it is due to, and thus increases with, the velocity gradient; that is, relative movements of objects within the material. 

\subsection{Link between discrete and continuous descriptions}
\label{link_discrete_continuous}

\subsubsection{Affine assumption}
\label{affine}

\begin{figure}
\centerline{\includegraphics[width=8cm]{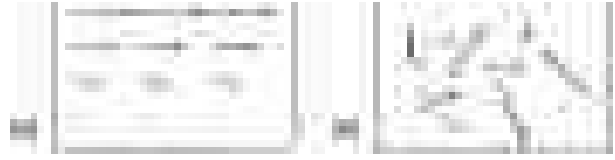}}
\caption{(a) Affine and (b) non affine flow field.}
\label{fig:AffineVersusNonAffine.eps}
\end{figure}

The {\em affine assumption}
 is analogous to, but much stronger than, eq. (\ref{velograd}).
It  assumes that the velocity of {\it each} individual object is affine too:
\begin{equation}
    \vec{v}\left( \vec{r}_2 \right)
    \stackrel{\rm affine}{\simeq}
    \vec{v}\left( \vec{r}_1\right)
    +
    \tensor{\nabla v}^t
    \cdot
     \left(   \vec{r}_2 - \vec{r}_1\right)
    .
    \label{v_affine_meso}
\end{equation}
An affine flow field and a non-affine flow field are plotted on figure \ref{fig:AffineVersusNonAffine.eps}.

In other words, this affine assumption implies that the continous velocity gradient has a meaning down to the level of individual objects, and that fluctuations around it are  small enough to have no effect on the material's mechanical behaviour.

In cellular patterns, especially in dry ones where there are no gaps nor overlaps, the movement of each individual object is highly correlated with its neighbours'; thus the affine assumption is reasonable
 \cite{Cartes2D}.
In particle assemblies, it might apply to dense assemblies of repelling particles, which cannot be too close nor too far from each other.

    Whenever this assumption is valid,
 it   considerably simplifies the description of the pattern evolution.
Consider for instance  a link, $\ellperso=\vec{r}_2-\vec{r}_1$ (eq. \ref{def_site_link}). Its time derivative is
$$
\frac{d\ellperso}{dt}
=  \vec{v}\left( \vec{r}_2 \right) -     \vec{v}\left( \vec{r}_1\right).$$
 Thus, under the affine assumption (eq. \ref{v_affine_meso}), the velocity gradient
modifies all links  in almost
the same way by (see eq. \ref{dem_link_under_affine}):
\begin{equation}
\frac{d\ellperso}{dt}
\;    \stackrel{\rm affine}{\simeq}\;
    \tensor{\nabla v}^t \cdot
 \;\ellperso.
 \label{link_under_affine}
 \end{equation}

In the definition (eq. \ref{eq:defC}bis) of $\tensor{C}^t$,
 the velocity gradient
can be taken out of the average:
\begin{eqnarray}
\tensor{C}^t&
    \stackrel{\rm affine}{\simeq}
&\left\langle\left(  \tensor{\nabla v}^t  \cdot \ellperso \right) \otimes\ellperso\right\rangle
\nonumber \\
&=&
 \tensor{\nabla v}^t\;\left\langle \ellperso\otimes\ellperso\right\rangle  .
\label{Ct_affine}
\end{eqnarray}
That is, 
\begin{equation}
\tensor{C} 
\;    \stackrel{\rm affine}{\simeq}\;
       \tensor{M} 
  \;\tensor{\nabla v}.
\label{C_affine}
\end{equation}
By injecting eq. (\ref{C_affine}) into eq.  (\ref{eq:defG}) we show that  $\tensor{W}$
is a statistical equivalent of the velocity gradient $ \tensor{\nabla v}$:
\begin{equation}
\tensor{W} =
    \tensor{M}^{-1} \tensor{C}
\;    \stackrel{\rm affine}{\simeq}\;
     \tensor{\nabla v}.
\label{G_affine}
\end{equation}
 This is why we included $\tensor{M}^{-1}$
only on the left side of $\tensor{W}$ (eq. \ref{eq:defG}).


\subsubsection{Velocity gradient and total strain rate}

By comparing eq. (\ref{deftotaldefo}) with eqs. (\ref{eq:defV},\ref{G_affine}) we identify
the statistical and dynamical definitions of
the total strain rate: \begin{equation}
\tensor{V}
   \stackrel{\rm affine}{\simeq}
\deps_{tot}.
\label{identific_dyn_stat}
\end{equation}

$\tensor{V}$ thus appears as  a statistical measurement of the symmetrised velocity gradient
 $\deps_{tot}$.
When only large scale measurements are possible, only  $\deps_{tot}$  can be measured.
However, when the detailed information on links in available to perform statistics,
measuring $\tensor{V}$  offers several advantages. \label{advantages_stat}

(i) The signal to noise ratio is optimal, in the sense that
all the local information, and only it, is used.
Each link acts as a small probe of the local velocity differences: the spatial derivative is taken naturally at the places where the objects are, not on the larger scale of RVEs. 

(ii)  $\tensor{{V}}$ is intrinsically based on the material's structure.
It can be defined and measured even if there are only a few objects; or if the standard deviation of their velocities is large.
Averaging over all links
provides a statistical measurement of the total strain rate.
At no point does the definition or measurement of $\tensor{{V}}$  require any affine description.

(iii) Physically, we expect $\tensor{{V}}$ to play a more general role than $\deps_{tot}
$, because it is based on the individual objects themselves. For instance, we expect $\tensor{{V}}$ to be determinant in yielding, and thus in the description of plasticity (and possibly $\tensor{\Omega}$ too)
 \cite{Cartes2D}. Similarly, the material's internal dissipations are probably more closely related to changes in the links than to a large scale velocity gradient: this suggests   that the dissipative contribution to the stress arises  in general from $\tensor{{V}}$ rather than from   $\deps_{tot}
$.


\subsubsection{Strain, in the elastic regime}

In this section we consider the particular case
where the material is in the elastic regime.
There is no plastic strain rate,
 $\deps_{pl}=0$.
Eq. (\ref{eq:total-strain-rate-is-shared}) becomes simply:
 \begin{equation}
\frac{{\cal D}\eps_{el}}{{\cal D}t}
\stackrel{\rm  elastic}{=}
\deps_{tot}
.
\label{noplastic} \end{equation}

 Thus, in the elastic regime, the elastic strain and the total  strain rate are not
independent physical quantities.
Combining eqs. (\ref{Landauluimeme}) and  (\ref{noplastic})
 shows that 
 \begin{equation}
\eps_{el}
\approx
\tensor{\eps}_{tot}.
\label{utile} \end{equation}
More precisely, at least in the linear elastic regime, one can identify two quantities:
the symmetrised gradient of the  displacement field, $\tensor{\eps}_{tot}$,
which is a function of the past history of the sample;
and  the elastic strain $\eps_{el}$,
which is a function of state.
In fact, in elasticity, both quantities are considered as equivalent
  \cite{lan86}.

   On the other hand, under the affine hypothesis,
  Ref. \cite{aub03} for the linear elastic regime (small strains),
and
Ref.   \cite{jan05}  for the non-linear elastic regime (large strains),
demonstrate that:
 \begin{equation}
\tensor{U} \;
  \stackrel{\rm affine}{\stackrel{\rm  elastic}{\simeq}}
\;
\eps_{tot}
.
\label{provedinaubouy}
 \end{equation}
 The demonstration of eq. (\ref{provedinaubouy})
  is similar to that for $\tensor{V}$
 (eqs.  \ref{Ct_affine}-\ref{identific_dyn_stat}):
it uses
the same hypotheses, with the additional assumption that   $\tensor{M}$ and $\tensor{M}_{0}$
commute (which is satisfied if $\tensor{M}_{0}$ is isotropic).

Eqs. (\ref{utile},\ref{provedinaubouy}) show that in the elastic regime:
   \begin{equation}
\tensor{U} \approx \eps_{el}.
\label{acceptable}
 \end{equation}
Thus the elastic
strain $\eps_{el}$ can be measured using two
different methods.  When large scale measurements of total
strain are possible, $\eps_{el}$ can be
measured as  $\eps_{tot}$.
 When the detailed information on links in available, measuring  $\eps_{el}$ as  $\tensor{U}$  offers many advantages, similar to that of  $\tensor{V}$
(section \ref{advantages_stat}).

An acceptable definition of strain must coincide with $\eps_{tot}$ in the linear elastic regime. 
As a consequence, it also implies that it is a conjugate of stress: the scalar product of stress by an infinitesimal increment of strain equals the increment of energy. This is a dynamical constraint on acceptable definitions; but it is a weak constraint (especially since the  conjugate equation is a scalar relation). In itself, it is insufficient to define all components of strain.

There are thus several families of acceptable definitions of  internal strain \cite{jan05,Bagi2006};
one family contains an infinity of acceptable definitions \cite{Farahani2000}.
 Some definitions are particularly adapted to a discrete pattern's geometry \cite{kru03} 
 or dynamics \cite{gol02}.

Here, 
eq.(\ref{eq:defU}) is a definition of strain which is: (i) one of the definitions acceptable  in the whole elastic regime, even at large strain (eq. \ref{acceptable}) where it coincides \cite{jan05} with a true strain  \cite{hog87};
(ii) probably the only definition valid outside of the elastic regime  \cite{jan05},
 when bubbles rearrange and move past each other, that is, when the pattern flows: the main advantage of eq. (\ref{eq:defU}) is that it does not require the detailed knowledge of each object's past displacement.     

\subsubsection{Plastic strain rate,  in steady flow }

In the more general case, there is a plastic strain rate,
  $\deps_{pl} \neq 0$, and deformations can be strongly non-affine.
 Eq. (\ref{noplastic}) does not hold.
The current elastic strain $\eps_{el}$  and the total strain rate $\eps_{tot}$ are
independent physical quantities;
 $\eps_{el}$ can no longer be measured as
 $\tensor{\deps}_{tot}$
 (whether it can be measured as $\tensor{U}$ is discussed
in section
 \ref{identify_general}).

For instance, if the material flows, the displacement of an object relatively to its neighbours can be arbitrary large;  $\deps_{pl} $ can become much larger than
 ${\cal D}\eps_{el} / {\cal D}t$.
In the  extreme cases of steady flows,  independent of time, it is possible (in absence of advection) that ${\cal D}\eps_{el}/{\cal D}t=0$,
and eq. (\ref{eq:total-strain-rate-is-shared}) reduces to:
\begin{equation}
 \deps_{tot}
 \stackrel{\rm  steady}{=}
  \deps_{pl}.
\end{equation}
According to eq.  (\ref{eq:defP}), a steady flow with a corotational derivative that vanishes (meaning no advection nor rotation effects, see eq. (\ref{eq:corotationnal})), implies that all  the geometrical strain rate translates into the topological strain rate:
$$
\tensor{P}\;
 \stackrel{\rm  steady}{=}
\; \tensor{V}.
$$
Using the identification of eq. (\ref{identific_dyn_stat}), we therefore obtain in that case:
\begin{equation}
\tensor{P}
\;
\stackrel{\rm affine}{\stackrel{\rm  steady}{\simeq}}
\;
\deps_{pl}.
\end{equation}

\subsubsection{Complete identification}
\label{identify_general}

The statistical tools $\tensor{U}$ and $\tensor{P}$
are always defined and measurable, even out of the elastic regime, or out of the steady regime.
If we could identify them with $\eps_{el}$ and
$\deps_{pl}$,  respectively,
it would make possible to measure the elastic strain in all regimes.
This is certainly not possible in general, as shown by both following counterexamples  \cite{referee}.


In granular systems, due to solid friction in the contacts, irreversible plastic strains appear before the list of contacts changes. In solid networks ({\it e.g.} solid foams) with no topological change, the bond themselves might behave plastically, or they might perhaps undergo buckling instabilities leading to non-reversible stress-strain curves. Those are examples in which plasticity occurs {\it before} the first topological change.

Conversely, consider a set of rigid cables which resist tension, but no compression, and tie them together at knots to form a redundant, hyperstatic network. Under given external forces on the knots, some cables will be taut, others will dangle and transmit no force. Upon changing the forces, the list of taut, tension-carrying cables will change. This can be regarded as a topological change. The response, which implies displacements and strains, is however reversible and might be called elastic. Hence a case for which plasticity begins {\it after} the first topological change.

This identification might turn possible in some particular cases where one can express the stress as a function of kinematical quantities. This seems to be the case for  foams and emulsions
 \cite{rau07,Cartes2D}.
We hope that in these cases, statistical measurements can constitute a coherent language to unify the description of elastic, plastic and fluid behaviours, as well as facilitate models and tests.


\section{Summary}

In the present paper, we define tools (Table \ref{table}) to
extract information from a
 pattern made of discrete objects,  subject to rearrangements,
 within
a wide class of complex
materials made of individual constituents  such as
 atoms, molecules, bubbles, droplets, cells or solid particles.
They characterise quantitatively the mutual arrangements of these objects,
or more precisely the links between neighbouring objects.

Their definition, which can flexibly adapt  to the questions to be answered, is operational. That is,
given an experimental
or simulated pattern, whether in 2D or 3D, there is a well defined method to measure them
directly as statistics on individual constituents (links between neighbouring
sites). This measurement is easy, and requires only a few basic operations
on a computer: multiplication, average, diagonalisation, logarithm.
It is  robust to experimental noise, even
if there is a limited number of links.

$\tensor{M}$, $\tensor{B}$ (or $\tensor{C}$) and $\tensor{T}$
characterise the current state of the pattern, its geometrical changes,
and its topological rearrangements, respectively.
They are explicitly based on
the pattern's discrete structure.
They can be measured locally, for
instance on a single biological cell, or grain in crystals.
But they can also be measured as averages over a larger region in
space, or as time averages. In foams, measuring them
smoothens out the pattern fluctuations due to the discrete nature of
bubbles, and evidences the underlying behaviour of the foam as a continuous
medium.

Their statistical counterparts
 $\tensor{U}$, $\tensor{V}$ (or $\tensor{W}$) and $\tensor{P}$,
 are independent of the pattern's discrete length scale.
 Each of them exists and is valid together
in elastic, plastic and fluid regimes : they unify the description
of these three mechanical behaviours.
They facilitate the comparison between experiments, simulations and theories.
In at least the linear, affine, elastic regime,
we suggest how to identify them with the quantities which characterise
the continous mechanics: elastic strain, total strain rate, and plastic strain rate,
respectively.
From a practical point of view, this offers the advantage of measuring these continuous quantities
with an optimal signal to noise ratio, even with few discrete objects.
On a fundamental side,
this provides a physical basis to the description of a continuous
medium, at any local or global scale, by relating it to the individual
constituents. Moreover, it provides a coherent language common to
elasticity, plasticity and fluid mechanics.

The companion paper \cite{Cartes2D} illustrates most of these points on a detailed practical example.

\section*{Acknowledgments}

This work was initially stimulated by a seminar delivered by G. Porte. We thank
M. Aubouy, S. Courty, J.A. Glazier, V. Grieneisen, M. Hindry, E. Janiaud,   Y. Jiang, J. K\"{a}fer, S. Mar\'ee for discussions.
We thank the colleagues who have made constructive comments about the first version of the manuscript.

\appendix


\section{Measurement techniques}

This Appendix, aimed at non-specialists, lists practical advices based on our past experience.

\subsection{Averaging procedure}
\label{average}

\subsubsection{Weights}

 The average
of any quantity $x$  is:  \begin{equation} \langle x\rangle=\frac{1}{N_{tot}}\sum w\; x,\end{equation}
where  the sum is taken
over all links in the averaging region. Here $w$ is the weight of the link: for almost all links  in the averaging region, $w=1$; at the boundaries of the averaging region, $w$ decreases to zero, different choices being possible (section
\ref{choice_weight}).
Here  we note: \begin{equation}N_{tot}=\sum w.\end{equation}
For instance, the texture is: \begin{equation}
\tensor{M}=\left\langle \tensor{m} \right\rangle =
\frac{\sum w\; \ellperso\otimes\ellperso}{\sum w}.
\label{def_M_poids}
\end{equation}

\subsubsection{Choices of weights}
\label{choice_weight}

\begin{figure}
\includegraphics[width=8cm]{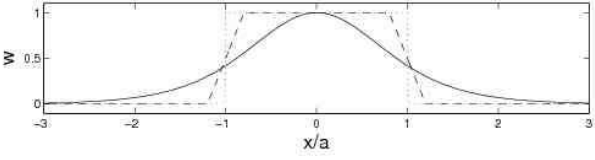}
\caption{Examples of averaging procedures: a link's weigth {\it vs} its position, here the box range is the segment $[-a,a]$. Dots:  "all or nothing". Dashes: "proportional". Solid line: "coarse grained", here with a hyperbolic tangent profile mirrored around the origin. }
\label{poids}
\end{figure}

There are at least three main possible choices for the averaging procedure (Fig. \ref{poids}). Once a procedure has been selected, it is important  to keep consistently the same for all measurements.

The {\it topology} is useful for local information, especially
for a single site: particle or cell. This is the case for instance when studying the division of a cell
\cite{courtypreprint}. In that case, each link is either included or excluded  ("all or nothing", $w=0$ or 1). One should at least include the links between the site of interest and its neighbours (first shell).
Statistics are over a few links only, and are easy to compute, even sometimes by hand.
This defines the $i$-th site's texture as a sum over its $n_i$ neighbours, labelled $j$: \begin{equation}
\tensor{M}_{i}
 =
 \frac{1}{n_i} \sum_{j=1}^{n_i}
\tensor{m}_{ij}
.
\label{eq:defM_local}
\end{equation}
One can also choose to include the second shell (next nearest neighbours), third, or even higher.
The total pattern's texture  (eq. \ref{def_M_poids}) appears as the average of the $n_{site}$ site textures, weighted by the site's number of links, and with a factor $1/2$ because each link is counted twice (each link belongs to two sites):
\begin{equation}
\tensor{M}
=  \frac{1}{2N_{tot}} \sum_{i=1}^{n_{site}} \sum_{j=1}^{n_i} \tensor{m}_{ij}
=\frac{1}{2N_{tot}}\sum_{i=1}^{n_{site}} n_{i}\tensor{M}_{i}.
\label{eq_Mtissue_moyen}
\end{equation}
The same definitions (eqs. \ref{eq:defM_local} and \ref{eq_Mtissue_moyen}) apply for $\tensor{B}$ or $\tensor{T}$.

The {\it geometry}  is useful for a continuous description, typically to measure  $\tensor{U} (\vec{R},t)$,
   $\tensor{P} (\vec{R},t)$  or $\tensor{V} (\vec{R},t)$ in a RVE as a function of space and time. This is the case for the examples of foam flow which illustrate this paper. The average is over all links in a
box which respects as much as possible the symmetry of the problem: rectangle, annulus.
The dimensions of the box determine the scale of averaging.
The measurements are performed on boxes at different positions $\vec{R}$. The distance between measurements positions cannot be more distant than the box size (it would leave gaps between boxes),
but they can be closer (thus boxes overlap).
If the box dimension is much larger than a link, one might choose to neglect  links which  cross  the box boundary.
But in general, links which cross the boundary require attention (especially near the box corner) if  an automatised
image analysis is used.
An additional choice is required.
 A first possibility ("all or nothing"), computationnally simpler, is to look where the link's center lies:
if the center lies inside the box, the link is assigned to this box ($w=1$); else, this link is not counted ($w=0$) \cite{asi03}. A variant is to assign half of each link  $(w=1/2)$  to the two boxes of the
two bubble centres it binds  \cite{dollet_local}.
A second possibility ("proportional"), which yields a better precision, consists in weighting the link with $w$ equal to the fraction  of the link which is inside the box (the remaining $1-w$ is outside)  \cite{jan05}.

The {\it coarse graining}   is seldom convenient in the practical applications considered here. However, theoreticians use it  \cite{gol02}, especially for the advection term (see Appendix \ref{time_evolution})
 \cite{dol05}.
A link at  position $\vec{r}$
is counted in a box at position $\vec{R}$ with a weight
$w(\vert\vec{r}-\vec{R}\vert)$. The coarse graining function $w$ is a function which is non-increasing,
from $w(0)=1$ to $w=0$, and has an integral equal to 1.  Its width at half height (that is, where $w=1/2$)
defines the scale of coarse graining.
It is a continous and differentiable function, so that advected links smoothly enter and leave  the averaging box, without singularity \cite{gol02}.

\subsection{ Choice of $\tensor{M}_0$ }
\label{M0_appendice}

Since  the reference texture $\tensor{M}_0$ plays almost no physical role, its choice is not very important.  It depends on the problem under consideration, but once its definition is chosen, it should be kept consistently. In practice, the choice depends on the available information. Here are  a few possibilities.

(i) The most favorable case is when $\tensor{M}_0$ can be measured directly.
In experiment, this is possible when an image can be chosen as reference, for instance a stress free pattern.
In simulation (Fig. \ref{patterns_particles}b), this requires to relax the stress under prescribed constraints.

(ii) $\tensor{M}_0$ can be determined theoretically in some cases, such as a set
 of particles which  interaction potential is known. This occurs in Fig. (\ref{patterns_particles}a), where the natural reference is the honeycomb pattern with a link size:
  \begin{equation}
    \ell_0^2
= 2\sqrt{3}A_{link}
= 2 \frac{A}{ \sqrt{3}}
        .
    \label{hexag}
\end{equation}
Here $A_{link}$ is
the area per link, and $\tensor{M}_0 = \ell_0^2 \tensor{I}_2/2$;
$A=3A_{link}$ is the area per particle  or cell  (section \ref{numbneighb}).

(iii)  In   cases  such as Fig. (\ref{patterns_foam}), no reference state is known in details.
 If only $\left \langle \ell_0^2 \right \rangle$ is known,
we suggest to take  $\tensor{M}_0$ as isotropic. Although we do
not know any fundamental reason for that, it seems to be
satisfactory in all practical cases we have encountered.
  From eqs.  (\ref{eq:Misotrope}, \ref{eq:Misotrope2D}) it writes, in $D=2$ or 3 dimensions:
  \begin{equation}
\tensor{M}_0 = \frac{\left \langle \ell_0^2 \right \rangle}{D} \tensor{I}_D
,
\label{eq:Misotrope_pourM0}
\end{equation}

(iv)  In some cases, $\left \langle \ell_0^2 \right \rangle$ is not known but we can estimate it. For instance, in a 2D cellular pattern of known average area
$\left\langle A \right\rangle $ (Fig. \ref{patterns_cells}), the comparison with hexagons (eq. \ref{hexag})
 suggests to take approximately :
 \begin{equation}
   \left \langle \ell_0^2 \right \rangle \approx \frac{2\left\langle A \right\rangle }{\sqrt{3}},
    \label{hexag_approx}
\end{equation}
 and $\tensor{M}_0 = \ell_0^2 \tensor{I}_2/2.$

(v) The most unfavorable case is when even $\left \langle \ell_0^2 \right \rangle$ is unknown.
A possibility is to take:
$$\tensor{M}_0 \approx  \bar{\lambda} \tensor{I}_D,$$
where $\bar{\lambda}$ is the average of the  $\lambda_i$s,
$\tensor{M}$'s eigenvalues. Taking the arithmetic average,
$\bar{\lambda}= \sum_i \lambda_i / D$, corresponds to the
assumption that $ \left \langle \ell^2 \right \rangle$ is
conserved: $\left \langle \ell_0^2 \right \rangle \approx \left
\langle \ell^2 \right \rangle$. Taking the geometric average,
$\bar{\lambda}= (\prod_i \lambda_i)^{1/ D}$, corresponds to the
assumption that Tr$\tensor{U}=0$, which is close to assuming that
the material is incompressible (see section \ref{ExU}).

\subsection{The case of 2D dry cellular patterns, especially foams}

\subsubsection{Number of neighbours}
\label{numbneighb}

In 2D dry cellular patterns, the number of neighbours of each cell is variable; but its average over the whole pattern is always close to 6 neighbours, and thus 6 links, per cell  \cite{wea99,weairerivier}. Since each link is shared by two cells, the number of links is 3 times the number of cells. This is also true for a moderately wet cellular pattern, if neighbours are defined on a skeletonized image. It  extends to Voronoi/Delaunay definition of neighbours for particles.

Some cells might meet by four ("4-fold vertex"). In that case, we recommend to decide that cells which share only a vertex should not be considered as neighbours. This choice is consistent with the fact that the texture describes the cell shape and arrangement (section \ref{inertia}). Moreover, this avoids many artefacts when measuring the T1s.

\subsubsection{Cell centers {\it versus} vertices}
 \label{vertex} 

Aubouy {\it et al.} \cite{aub03} chose to describe a cellular pattern (such a 2D dry foam) as a network, each site being a vertex (that is, a point where three cells meet).
Here, we  prefer to use cell centers, for several reasons.

(i)
First, and most important: the centers move according to the overall velocity field (while vertices have a highly fluctating displacement), thus the affine assumption
(eq.  \ref{v_affine_meso}) applies.

(ii)
It is more robust, because a cell center is measured as an average over several pixels (while a vertex is a single pixel, which position might depend on the image analysis procedure).

(iii)
 This has the advantage of being more general: it applies to all other discrete patterns; and even within cellular patterns, it generalises to wet foams, and to 3D.

(iv) Finally, the topological rearrangements are well characterised (while on the opposite, if $\ell$ was defined as the vector between two vertices, a T1 occurence would be defined as $\ell_a=\ell_d=0$, so that the contribution of a T1 to eq. (\ref{eq:defT}) would systematically be zero).

\subsubsection{Texture and inertia matrices}
\label{inertia} 
Mar\'ee {\it et al.} \cite{mar07}
propose to measure the shear modulus by considering the variation of cell shapes.
Each cell's shape is characterised by its inertia matrix, $ \left \langle \vec{r} \otimes \vec{r} \right \rangle$: it looks similar to
eq. (\ref{eq:brut_lisible3D}), but it is averaged over the position $\vec{r}$ of pixels   inside a cell;
thus their description is {\it intra-cellular}. Ours,
averaged over the links  between the cells, and thus
 based on the shape of the overall pattern, is rather {\it inter-cellular}.

In dry cellular patterns, where there are no gaps between cells, nor overlaps, the deformation of each cell is highly correlated to the global strain; thus, in this case, both descriptions coincide
and yield approximately the same results \cite{courtypreprint}.

For instance, note that the shear modulus
\label{shear_modulus} 
 is the variation of elastic stress with respect to infinitesimal variations of $ \tensor{U}$. This measurement  is robust
\cite{asi03,jan05}.
As mentioned in section  \ref{prefactor}, it is not affected if we multiply
$\tensor{M}$ and $\tensor{M}_0$ by a same prefactor; and even if we change $\tensor{M}_0$, see for instance eq. (\ref{eq:defU}).
This is why  this particular measurement gives similar results with both inertia and texture.

Here, we  prefer to use the texture based on cell centers, which is more general, for several reasons.

(i)  It also applies to characterise the strain of wet foams
(where bubbles are round, and thus each bubble's inertia is isotropic).

(ii) It applies to all other discrete patterns, including particle assemblies.

(iii) Centers, rather than shape, are involved in the kinematic description, including eqs.
(\ref{eq:defV},\ref{eq:defP}). It could in principle be possible to define an equivalent of  $\tensor{B}$ (and even of $\tensor{V}$) based on inertia matrix, but its physical meaning is unclear; and it is probably not possible to define an equivalent of $\tensor{T}$ (and  $\tensor{P}$).

(iv) It extends to more than one cell; while the inertia matrix of several cells can be defined, its physical meaning is not relevant to the pattern description.

Note that in  the graphical representation of the inertia matrix, the ellipse axis lengths are the square root of the matrix' eigenvalues  \cite{mar07}.  The advantage is that the ellipse elongation is the same as that of the actual cell. Here, taking the square root of eigenvalues has no physical signification for any matrix (except for the texture), so that we plot the matrix' eigenvalues themselves (section \ref{meter}).
 

\section{Matrices: notations and definitions}
\label{notation_tenseurs}

This appendix is aimed at readers who are not familiar with the matrices. We list all standard definitions used in the text, from the simplest to the most complicated.

 \subsection{Matrices}
 \label{not_tens}

We work here in a space with $D=3$ dimensions. A  {\it scalar} is a simple number; a {\it vector} is a list of $D$ numbers; a {\it matrix} is an array
of $D\times D$ numbers. All these objects are {\it tensors}, of rank 0, 1 and 2, respectively.
In this paper, there also appears
 ${\cal J}$, which is a
tensor of rank 3 (for which there exists no particular name), that is, an array
of $D^3$ numbers.



 A matrix  $\tensor{A}$ is an array with components $A_{ij}$, where the indices $i,j=1$, 2 or 3:
  \begin{equation}
 \tensor{A} = \left(
\begin{array}{ccc}
A_{11}&A_{12}&A_{13}\\
A_{21}&A_{22}&A_{23}\\
A_{31}&A_{32}&A_{33}
 \\  \end{array}
\right)
.
\label{array_detail}
\end{equation}
Its {\it trace} is the sum of its diagonal terms:
\begin{equation}
{\rm Tr} \; \tensor{A} = A_{11} + A_{22} + A_{33}.
\label{trace}
\end{equation}
Its {\it transposed}  $\tensor{A}^t$  has  components $A^t_{ij}= A_{ji}$, and has the same trace.
Any matrix can be rewritten as the sum of its {\it symmetric}   and  {\it antisymmetric } parts:
$$
\tensor{A} \; =\;     \frac{\tensor{A} + \tensor{A}^t}{2} \; + \; \frac{\tensor{A} - \tensor{A}^t}{2}.
$$
A matrix $\tensor{S}$ is said to be symmetric if it is equal to its transposed, $\tensor{S} = \tensor{S}^t$, that is, $S_{ij}=S_{ji}$
(while an antisymmetric matrix is equal to minus its transposed);
by definition, the symmetric part of $\tensor{A}$ is always symmetric.
A symmetric matrix can itself be rewritten as an isotropic term and a traceless (or {\it deviatoric}) term:
 \begin{equation}
\tensor{S} \; =\;   \frac{{\rm Tr}( \tensor{S})}{D} \; \tensor{I}_D + {\rm Dev}( \tensor{S}),
\end{equation}
where $\tensor{I}_D$ is the identity matrix in dimension $D$
(appearing in eq. \ref{eq:Misotrope}). 

Itself, the deviatoric part can be decomposed in diagonal components, called {\it normal differences}, and off-diagonal ones.

To summarize, a matrix has in general  9 independent components $A_{ij}$. They can be rewritten as
3 antisymmetric ones, namely $(A_{12}-A_{21})/2$, $(A_{23}-A_{32})/2$, and $(A_{31}-A_{13})/2$;
and 6 symmetric ones, namely 1 trace $A_{11} + A_{22} + A_{33}$,
2 normal differences $A_{11}- A_{22}$ and  $A_{22} - A_{33}$, 3 off-diagonal terms
 $(A_{12}+A_{21})/2$, $(A_{23}+A_{32})/2$, and $(A_{31}+A_{13})/2$.
This means that an antisymmetric matrix has 3 independent components,  a symmetric matrix one 6, a deviatoric one has 5, an isotropic one has 1.

 The product between matrices is another matrix:
 \begin{equation}\left(\tensor{A}\tensor{B}\right)_{ij} = \sum_{k}A_{ik}B_{kj}.  \end{equation}
The product between a matrix and a vector is another vector:
 \begin{equation}\left(\tensor{A}\cdot \vec{a}\right)_{i} = \sum_{k}A_{ik}a_{k}.  \end{equation}
The scalar product between matrices is a number:
 \begin{equation}\tensor{A}:\tensor{B} = \sum_{i,k}A_{ik}B_{ki} = {\rm Tr}
 \left(\tensor{A}\tensor{B}^t\right).  \end{equation}
The ("euclidian") {\it norm} of $\tensor{A}$ is a strictly positive  number defined in any dimension $D$ as:
 \begin{equation}
\vert \vert  \tensor{A}\vert \vert = \sqrt{\tensor{A}:\tensor{A}} = \left[ \sum_{i,k}(A_{ik}A_{ki})\right]^{1/2}.
 \label{def_norm}
 \end{equation}

Many practical applications regard 2D images.
 Matricial notations are valid in any dimension $D$, and it is straigthforward
 to rewrite them in 2D, see also section  \ref{2Dcase}.
   In 2D, a matrix $\tensor{A}$
 has in general 4 independent components $A_{ij}$, where $i,j=1$ or 2. They can be rewritten as
1 antisymmetric one, namely $(A_{12}-A_{21})/2$;
and 3 symmetric ones, namely 1 trace $A_{11} + A_{22}$,
1 normal difference $A_{11}- A_{22}$, 1 off-diagonal terms
 $(A_{12}+A_{21})/2$.
This means that an antisymmetric matrix has 1 independent component,  a symmetric matrix has 3, a deviatoric one has 2, an isotropic one has 1.

 \subsection{Diagonalisation}
 \label{diagonale}

For a  symmetric matrix $\tensor{S}$,
there   exist three orthogonal axes,
called $\tensor{S}$'s {\it eigenvectors} (from the German word "eigen", meaning "own"), in which  $\tensor{S}$ would be diagonal, see for instance eq. (\ref{eq:Mdiagonal}). That is, if we used these axes (instead of the original ones) to measure the matrix, it would have
non-zero terms only along its diagonal:
\begin{equation}
 \tensor{R} \tensor{S} \tensor{R}^{-1} = {\rm diag} \;  \tensor{S} =
\left(
\begin{array}{ccc}
{s}_1 &0&0\\
0&{s}_2&0\\
0&0&{s}_3
\\  \end{array}
\right)
.
\label{eq:Sdiagonal}
\end{equation}
Here $ \tensor{R}$ is the matrix of rotation from the original axes to the eigenvectors. 
The three numbers ${s}_1$, ${s}_2$, ${s}_3$ are called the matrix' {\it eigenvalues}. We label them in order of decreasing absolute value:  $\vert {s}_1 \vert \ge \vert {s}_2 \vert \ge \vert {s}_3 \vert$.

 They determine many properties of $\tensor{S}$, including its trace and norm:
\begin{eqnarray}
{\rm Tr}( \tensor{S}) &=& {s}_1 + {s}_2+ {s}_3
,\nonumber \\
\vert \vert  \tensor{S}\vert \vert  &=&
 \sqrt{ {s}_1^2 + {s}_2^2+ {s}_3^2}.
\end{eqnarray}
If they are non zero,
the inverse of $\tensor{S}$ exists, and it is diagonal in the same axes as  $\tensor{S}$:
\begin{equation}
{\rm diag} \;  \tensor{S}^{-1} =
 \tensor{R}^{-1}
 \left(
\begin{array}{ccc}
\frac{1}{{s}_1} &0&0\\
0&\frac{1}{{s}_2}&0\\
0&0&\frac{1}{{s}_3}
\\  \end{array}
\right)
 \tensor{R}
.
\label{eq:Sinverse}
\end{equation}
If they are strictly positive, the {\it logarithm} of $\tensor{S}$
(see eq. \ref{deflogM}) is defined by rotating
to the eigenvectors, taking the logarithm of the eigenvalue, and rotate back to the original axes:
\begin{equation}
\log  \tensor{S}=  \tensor{R}^{-1}
   \left(
\begin{array}{ccc}
\log{{s}_1} &0&0\\
0&\log{{s}_2}&0\\
0&0&\log{{s}_3}
\\  \end{array}
\right)
 \tensor{R}
.
\label{eq:Slog}
\end{equation}
By construction, log$\tensor{S}$ is symmetric too, and diagonal
in the same axes as $\tensor{S}$.

The literature of mechanics
 \cite{cha87,fra95} sometimes uses a specific definition of shear.
 It is characterised by a deviatoric matrix with two opposite eigenvalues (${s}_1=-{s}_2=S$) and nothing in the third direction (${s}_3=0$). Its {\it amplitude} $S$ is defined as:
 \begin{equation}
 S =   \left[\frac{1}{2}\; \sum_{i,j} S_{ij}^2\right]^{1/2} = \frac{ \vert \vert  \tensor{S}\vert \vert }{\sqrt{2}}.
 \label{def_amplitude}
 \end{equation}

\begin{figure}
\includegraphics[width=6cm]{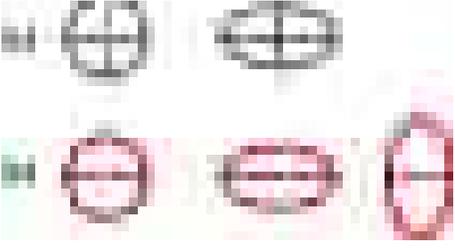}
\caption{Ellipses to represent matrices of eigenvalues $s_{1}$ and $s_{2}$, with a solid line to represent positive eigenvalues. (a) Two positive eigenvalues are represented by a "crossed ellipse". The circle represents an isotropic matrix, $s_{1}=s_{2}$. (b) When there is one positive and one negative eigenvalue, 
the circle represents $s_{1}= - s_{2} >0 $; the "coffee bean" ellipse is elongated along the positive eigenvalue ($s_{1}> - s_{2}> 0$);  the "capsule" ellipse  is elongated along the negative eigenvalue
($-s_{1}>  s_{2}> 0$). }
\label{fig:EspecesEllipses.eps}
\end{figure}

In 2D,  a matrix can be represented graphically by an ellipse, which axes, represented by solid lines, are in the directions $\theta$ and $\theta+90^\circ$, and have a length $s_1$ and $s_2$, respectively; the sign of the eigenvalues is labeled specifically by plotting  a line for a positive eigenvalue, and no line for a negative eigenvalue (Fig. \ref{fig:EspecesEllipses.eps}). 
The ellipse anisotropy   $\eta= (\vert s_1 \vert -  \vert s_2\vert )   / \vert s_1 /\vert  = 1 - \vert s_2   / s_1 \vert $ is between 0, for a circle, and 1, for an extremely thin ellipse. The ellipse size is characterised by $\vert s_1 \vert +  \vert s_2\vert$. 

If $\tensor{S} $ has strictly positive eigenvalues,
 $s_1 \ge s_2  >0$, it  is entirely defined by three numbers: first,  its trace  ${\rm Tr}(\tensor{S}) = s_1 + s_2 $, equal to the ellipse's characteristic size; second, its anisotropy  $\eta=   1 - s_2   / s_1 $, equal to that of the ellipse;  third,
 the direction  $\theta$ of its largest eigenvalue's axis  ($0^{\circ}\leq\theta<180^{\circ}$), which is ill-defined for an isotropic matrix ($\eta$ close to 0). Its determinant $ s_1s_2 $ is proportional to the ellipse area, but  it is not used in the present paper.

 \subsection{Outer product}
\label{outerprodu}

The {\it outer product} (or  {\it tensor product}) of two vectors $\vec{a}$, $\vec{b}$ is the matrix
of components $a_ib_j$:
  \begin{equation}
 \vec{a} \otimes \vec{b}
 = \left(
\begin{array}{ccc}
a_1b_1&a_1b_2&a_1b_3\\
a_2b_1&a_2b_2&a_2b_3\\
a_3b_1&a_3b_2&a_3b_3
 \\  \end{array}
\right)
.
\label{outer}
\end{equation}
Its trace is the scalar product $\vec{a} \cdot \vec{b} $:
$${\rm Tr} ( \vec{a} \otimes \vec{b} ) = \vec{a} \cdot \vec{b}= a_1b_1 + a_2b_2 + a_3b_3.$$
For instance, eq. (\ref{eq:brut_lisible3D}), and eq. (\ref{eq:defM_lisible3D}) or  (\ref{eq:defM_lisible2D}), write:
\begin{eqnarray}
\tensor{m}&\equiv & \ellperso\otimes\ellperso
, \nonumber \\
\tensor{M}&\equiv&\left\langle \ellperso\otimes\ellperso\right\rangle
.
\label{eq:defM_compacte}
\end{eqnarray}
In eq. (\ref{eq:defM_compacte}),   the physical interpretation of the outer product  by $\ellperso$ is that it  transforms the surface integral of a
discrete individual vector, the link, into a bulk integral of a continuous average matrix, the texture  \cite{aub03,jan05}.
\label{stress}

 The outer product is also used  for the notation
$\vec{\nabla}=(\partial / \partial  r_i) = (\partial/\partial x, \partial/\partial y, \partial/\partial z)$
 (called gradient, or {\it nabla}) which symbolises the space derivatives:
 \begin{eqnarray}
 \tensor{\nabla v}
&=&
\vec{\nabla}\otimes\vec{v}
= \frac{ \partial v_j}{ \partial r_i}
\nonumber \\
&=&
 \left(
   \begin{array}{ccc}
 \frac{ \partial v_1}{ \partial x}& \frac{ \partial v_2}{ \partial x}& \frac{ \partial v_3}{ \partial x}\\
 \frac{ \partial v_1}{ \partial y}& \frac{ \partial v_2}{ \partial y}& \frac{ \partial v_3}{ \partial y}\\
 \frac{ \partial v_1}{ \partial z}& \frac{ \partial v_2}{ \partial z}& \frac{ \partial v_3}{ \partial z}\\
\end{array}
\right)
\label{defgradv}
\end{eqnarray}
More precisely, eq. (\ref{velograd}) uses its transposed:
   \begin{equation}
    \tensor{\nabla v}^t= \frac{ \partial v_i }{ \partial r_j}.
\label{defgradv_transposed}
   \end{equation}
Similarly, the rotational is the vector product $\nabla_i \times v_j$.

However, rheologists \cite{cha87,fra95} often prefers the notation
${\rm grad}v=  \partial v_i  / \partial r_j.$ This creates an ambiguity with eqs.
(\ref{defgradv},\ref{defgradv_transposed}). In case of doubt, it is safe to come back to indices, which are
unambiguous.
For instance, the demonstration of eq. (\ref{link_under_affine}) writes as follows:
\begin{equation}
   \frac{\partial \ell_i}{\partial t} = v_i (r_j+\ell_j)-v_i(r_j) = \sum_j \frac{\partial v_i}{\partial r_j}\ell_j .
\label{dem_link_under_affine}
  \end{equation}

In section \ref{sec:defJ} and in eq. (\ref{divided}),  the notation ${\cal J} =  \vec{v} \otimes \tensor{M}$
means $ {\cal J}_{ijk} = v_iM_{jk}$, and the notation
$\vec{\nabla} \cdot {\cal J}$ is equivalent to $\sum_i \partial {\cal J}_{ijk}/\partial r_i.$

\section{Time evolution}

\subsection{Finite time interval $\Delta t$}
\label{time_evolution}

This appendix can be useful for a user who wants to analyse a movie (and not only a static image).
It helps to understand the definitions, units and measurements of $\tensor{B}$ and $\tensor{T}$; as well as
the time evolution of  $\tensor{M}$ (eq. \ref{time_evolution_M}).

 \subsubsection{Time interval between images}

We consider a movie.
To obtain good measurements,
 averages can be performed
on a large time interval $\tau$, that is, a large number $\tau/\Delta t$ of images.
Here $\Delta t$ is the time interval between  two consecutive images, at times $t$ and $t+\Delta t$.
Ideally, measurements should not depend too much on the exact value of  $\Delta t$.
In practice, however small $\Delta t$ is, it is finite, and this discretisation has consequences, see below
eqs. (\ref{correction_moyenne},\ref{terme_carre})

If possible, $\Delta t$ should be chosen small enough to enable a good tracking of objects from one image to the next (typically, during $\Delta t$, relative displacements of sites should be a fraction of the inter-sites distance). This is unimportant if objects are labelled individually, as is the case in simulations.

$\Delta t$ should also be chosen small enough that for any time-dependent (but space-independent) quantity $x$,
$\Delta x/\Delta t$ tends towards its time derivative
$dx/dt$.
When $x$ depends both on space and time, $\Delta x/\Delta t$ tends towards
$dx/dt$  if the measurement box moves along with the links (so-called "Lagrangian" point of view,  often useful in theory). When the   measurements are performed in a fixed region of space (so-called "Eulerian" point of view, often useful in practice, especially in steady flows),
  $\Delta x/\Delta t$ tends towards its partial time derivative $\partial x/\partial t$.

Eq. (\ref{def_M_poids}) can be rewritten as:
 \begin{equation}
N_{tot}\tensor{M}=\sum w\; \ellperso\otimes\ellperso =\sum w\; \tensor{m}.
\label{a_varier}
\end{equation}
Its variation between successive images,
divided by $\Delta t$, involves the links $\ellperso_{a}$ (resp. $\ellperso_{d}$) appeared (resp. disappeared) during $\Delta t$.
This means that $N_{tot}$ and $\tensor{M}$ are evaluated both at $t$ and $t+\Delta t$;
 $\ellperso_{a}$ and $\tensor{m}_a$ (resp. $\ellperso_{d}$ and $\tensor{m}_d$)  are evaluated 
at $t+\Delta t$ (resp. $t$).
Then:
\begin{eqnarray}
\Delta  \tensor{m} 
&=&
  \Delta \left( \ellperso\otimes\ellperso\right)
\nonumber \\
&=&
      \Delta \ellperso \otimes\ellperso + \ellperso\otimes     \Delta  \ellperso  
+ \xi    \Delta   \ellperso  \otimes     \Delta  \ellperso  
.
\label{correction_moyenne}
\end{eqnarray}
Here $\xi=1$ if $\ellperso $ is evaluated at time $t$; $\xi=-1$ if $\ellperso $ is evaluated at time $t+\Delta t$;
and  $\xi=0$ if $\ellperso $ is the average of its values at $t$ and $t+\Delta t$, which is recommended
to simplify eq. (\ref{terme_carre}). 

\subsubsection{Effect of discrete time }

The variation of eq. (\ref{a_varier}) between successive images writes:
 \begin{eqnarray}
\frac{\Delta (N_{tot}\tensor{M})}{\Delta t} & = & \sum\frac{\Delta w}{\Delta t}\tensor{m}
 + \sum w
 \frac{\Delta \tensor{m}}{\Delta t}   \nonumber \\
 && 
+\frac{
\sum_{a} w_a\tensor{m}_{a}  -\sum_{d} w_d \tensor{m}_{d} 
}{\Delta t}
.
  \nonumber\\
  \label{complete_oeuf_jambon_fromage}
  \end{eqnarray}

Quantities in eq.   (\ref{complete_oeuf_jambon_fromage}) are extensive, and are convenient for practical measurements. However, for theory, intensive quantities are easier  to manipulate (Table \ref{table}).
Dividing both sides by $N_{tot}$ yields:
\begin{equation}
\frac{\Delta\tensor{M}}{\Delta t}+\frac{\tensor{M}}{N_{tot}}\frac{\Delta  N_{tot}}{\Delta t}
=-\vec{\nabla} \cdot {\cal J}_{M}
 +\tensor{B}+\tensor{T}.
\label{divided}
\end{equation}
Eq. (\ref{divided}) tends towards  eq. (\ref{time_evolution_M}) in the limit of small $\Delta t$, as we now show term by term.


The relative variation of $N_{tot}$ during $\Delta t$ is  $\Delta\log N_{tot}/\Delta t$, and at small $\Delta t$
it tends towards $\partial \log N_{tot}/\partial  t$.

The advection term
$-\vec{\nabla} \cdot {\cal J}_{M}$ is the
term in  $\Delta w/\Delta t $, in  eq. (\ref{complete_oeuf_jambon_fromage}).
It is due to links entering or exiting the
region where $\tensor{M}$ is measured (see section \ref{sec:defJ}).
In a first approximation,  ${\cal J}_{M}\simeq \vec{v} \otimes \tensor{M}$. The demonstration   is delicate and we do not develop it here. Briefly, when the averaging procedure uses a  coarse-graining function
$w(\vec{r}(t))$, it is possible to transform
  a time derivative of $w$ into a space derivative; this involves $d\vec{r}/dt$, that is, the local velocity
  \cite{dol05,gol02}.

The geometrical variation term is:
\begin{equation}
\tensor{B}=\frac{N_{c}}{N_{tot}}
\left\langle
 \frac{\Delta \tensor{m}}{\Delta t}
 \right\rangle .
\label{B_discrete}
\end{equation}
Here $N_{c}$ is the number of links conserved between both images, that is, the number of
terms involved in the average noted $\left\langle . \right\rangle$.
Most links contribute to eq. (\ref{divided}), but each one has a small contribution.
If $\Delta t$ is small enough, then $\Delta\ellperso/\Delta t$
tends towards $d\ellperso/dt$ and the correction $N_{c}/N_{tot}$ tends towards
1, so that we obtain eq. (\ref{eq:defB}).
Eq. (\ref{correction_moyenne}), 
implies that:
\begin{equation}
\tensor{B}= \tensor{C}+\tensor{C}^{t} + {\cal O}(\xi),
\label{terme_carre}
\end{equation}
where $\tensor{C}$ is defined as:
\begin{equation}
\tensor{C}=\frac{N_{c}}{N_{tot}}\left\langle
\ellperso
\otimes
\frac{\Delta\ellperso}{\Delta t}
\right\rangle .
\label{defCbrute}
\end{equation}
If we choose  $\xi=0$, eq. (\ref{terme_carre}) means that 
 $\tensor{B}$ is twice the symmetrical part of $\tensor{C}$:
 \begin{equation}
\tensor{B}= \tensor{C}+\tensor{C}^{t} + {\cal O}(\xi),
\label{terme_sans_carre}
\end{equation}
 If $\Delta t$ is small enough,  eq. (\ref{defCbrute}) yields:
 \begin{eqnarray}
\tensor{C}&=&\left\langle
\ellperso
\otimes
\frac{d\ellperso}{dt}
\right\rangle
\nonumber \\
&=&
\left(
\begin{array}{ccc}
 \left \langle X\frac{dX}{dt}\right \rangle & \left \langle Y\frac{dX}{dt} \right \rangle & \left \langle Z\frac{dX}{dt} \right \rangle \\
 \left \langle X\frac{dY}{dt} \right \rangle & \left \langle Y\frac{dY}{dt} \right \rangle & \left \langle Z\frac{dY}{dt} \right \rangle \\
 \left \langle X\frac{dZ}{dt} \right \rangle & \left \langle Y\frac{dZ}{dt} \right \rangle & \left \langle Z\frac{dZ}{dt} \right \rangle \\
 \end{array}
\right)
.
\label{eq:defC}
\end{eqnarray}
In general, $\tensor{C}$ is not symmetric; its transposed
is $\tensor{C}^{t}=\left\langle  d\ellperso/dt \otimes \ellperso\right\rangle$.

The topological term is between the last parentheses in eq.
(\ref{complete_oeuf_jambon_fromage}).
Note that we can  derive its exact prefactor:
 \begin{equation}
\tensor{T}=\frac{1}{\Delta t}\;\frac{\Delta N_{a}}{N_{tot}}\;\left\langle \tensor{m}_{a}\right\rangle -\frac{1}{\Delta t}\;\frac{\Delta N_{d}}{N_{tot}}\;\left\langle  \tensor{m}_{d}\right\rangle ,
\label{T_complete}
\end{equation}
 where $\Delta N_{d}$ is
the number of disappeared links (that exist at $t$ but no longer
at $t+\Delta t$); and $\Delta N_{a}$ the number of appeared
links (that exist at $t+\Delta t$ but not yet at $t$).
In the limit of small $\Delta t$, eq. (\ref{T_complete}) tends towards  eq. (\ref{eq:defT}).
In eq. (\ref{T_complete}), the number of terms involved in the average
$\left\langle \tensor{m}_{a}\right\rangle $
(resp: $\left\langle \tensor{m}_{d}\right\rangle $) is $\Delta N_{a}$
(resp: $\Delta N_{d}$), which is much smaller than $N_{tot}$.
This represents a small number of links, each having a large contribution to eq. (\ref{divided}).
Thus statistics on
$\tensor{T}$ are always much noisier than that on $\tensor{M}$ or $\tensor{B}$,
and it is advisable to integrate $\tensor{T}$ over a long time $\tau$.

\subsection{Objective time derivatives}

This technical section is rather aimed at specialists. It discusses the different
objective derivatives which appear in the course of this paper, when estimating
the time derivatives of $\tensor{M}$ or $\tensor{U}$. By definition, the objective derivative 
of a matrix is invariant after a change in any other rotating frame of reference. 
It expresses the fact that  $\tensor{M}$ or $\tensor{U}$ are intrinsic properties of the material.
In practice the objective derivative writes as the sum of a total (Lagrangian) derivative 
plus corrections involving the velocity gradient.
In principle there is an infinity of possible objective derivatives.
The following sections show that objective derivatives 
can be selected and calculated for the evolution of  matrices $\tensor{M}$ and $\tensor{U}$.

\subsubsection{Time evolution of $\tensor{M}$}

In the affine assumption, the geometrical term $\tensor{B}$ of
eq. (\ref{time_evolution_M})
can be rewritten  using eqs. (\ref{eq:defG},\ref{G_affine},\ref{terme_sans_carre}):
\begin{eqnarray}
\tensor{B}
&=&
   \tensor{M}  \; \tensor{W}
    +  \tensor{W}^{t} \; \tensor{M}
    \nonumber \\
&   \stackrel{\rm affine}{\simeq}&
     \tensor{M} \; \tensor{\nabla v}
    + \tensor{\nabla v}^{t} \; \tensor{M}.
\label{convectiveterm}
\end{eqnarray}

Thus, the second term of the right hand side of
eq.  (\ref{time_evolution_M})
can be
grouped with its left hand side, formally appearing as a Maxwell
upper convective tensor derivative (see e.g.~\cite{mac94}):
\begin{equation}
    \stackrel{\nabla}{\tensor{M}}
    \ =  \
      \frac{\partial \tensor{M}}{\partial t}
    + \vec{v} \cdot \nabla \tensor{M}
    - \tensor{\nabla v}^{t}\; \tensor{M}
    - \tensor{M}\; \tensor{\nabla v}.
    \label{derivee_Maxwell}
\end{equation}
Here we have assumed incompressibility to transform $\vec{\nabla} \cdot {\cal J}_{M}
\approx \vec{\nabla} \cdot \left(\vec{v} \otimes \tensor{M} \right) $
into
$\vec{v} \cdot \vec{\nabla} \tensor{M} $.
Hence
eq.  (\ref{time_evolution_M})
 appears as a conservation
equation for $\tensor{M}$:
\begin{equation}
    \stackrel{\nabla}{\tensor{M}}
    \ \ \stackrel{\rm affine}{\simeq} \
    - \tensor{{T}},
    \label{derivee_M_affine}
\end{equation}
which source $\tensor{{T}}$
is due to topological changes.
Note that the present approach has unambiguously selected the {\it upper}
(rather than the lower, or any other) convective
tensor derivative.

\subsubsection{The small $\tensor{U}$  assumption}
 \label{upperconvectedderivative}


Inverting eq. (\ref{eq:defU}), the texture develops as:
\begin{equation}
\tensor{M}  =    \tensor{M_0} \exp \left(2\tensor{U}\right)
 .
\label{inverti}
\end{equation}
 In a plastic material such as considered here,
the elastic internal strain is seldom much larger than unity.
This is the case for foams, where deformation of bubbles does not excess the size of two bubbles before topological changes
occurs, and thus $\vert \vert \tensor{U} \vert \vert $ is bounded.
In the case where the strain $\tensor{U}$ is small everywhere,
eq. (\ref{inverti}) becomes:
\begin{equation}
   \tensor{M}
    =  \tensor{M_0}  \left[  \tensor{I} + 2\tensor{U}  +
      {\cal O}\left( \tensor{U} ^2\right)   \right]
   \stackrel{\rm small}{\simeq}   \tensor{M_0} +    2    \tensor{M_0}    \ \tensor{U}.
   \label{a_imiter}
\end{equation}
   Let us also assume that the reference configuration is isotropic,
   $\tensor{M}_0=M_0 \tensor{I}$.
Then, eq. (\ref{derivee_M_affine}) becomes:
\[    2 M_0    \left(  
\stackrel{\nabla}{\tensor{U}}      
   -
\frac{  \tensor{\nabla v}  + \tensor{\nabla v}^{t}}{2}
    \right)
    \ \stackrel{\rm small}{\simeq} \     - \tensor{T}, \]
or equivalently:
\begin{equation}
 \stackrel{\nabla}{\tensor{U}} 
 \ \stackrel{\rm small}{\simeq} \
 \tensor{V}
       - \tensor{P},
    \label{plasticaffine}
 \end{equation}
where $\tensor{P} =  - \tensor{T}/(2 M_0)$  is  the rate of the plastic strain.
This linear assumption thus simplifies
eqs.  (\ref{eq:Vshared},\ref{eq:corotationnal}).
Again, this selects an {\it upper} convective tensor derivative.

\subsubsection{Kinematic equation of $\tensor{U}$}
\label{kine_eq_obj} 

 In section \ref{upperconvectedderivative}, with
 $\tensor{U}$   small and $\tensor{M}_0$ isotropic,
$\tensor{M}$   commuted with
 its time derivative. Thus,
 knowing the time derivative of
 $\tensor{M}$,
  eq. (\ref{eq:defU}) immediately yielded the time derivative of $\tensor{U}$.
 This enabled to eliminate  $\tensor{M}$ from the time evolution of  $\tensor{U}$
 (eq.   \ref{plasticaffine}).


This is not the case in general.
 If the eigenvectors of $\tensor{M}$ change (rotate) with time,
 $\tensor{M}$ does not necessarily commute with
 its time derivative.
There is no simple relation between the time derivatives of
 $\tensor{M}$ and $\tensor{U}$.

It is thus tedious to obtain the time evolution of  $\tensor{U}$ {\it versus}
$\tensor{U}$, instead of  {\it versus} $\tensor{M}$.
We do not develop here the calculation.
Briefly,  we
start from eq. (\ref{a_imiter}). We differentiate it:
  \begin{equation}
\frac{\partial\tensor{M}}{\partial t}
    =  \tensor{M_0}  \left[  \tensor{I} + 2\frac{\partial\tensor{U}}{\partial t}  +
      {\cal O}\left( \tensor{U} ^2\right)   \right]
.
\label{ca_approxime}
\end{equation}
We then
 inject eq. (\ref{ca_approxime}) in the time evolution of $\tensor{M}$ (eq. \ref{time_evolution_M}) at lowest order terms in $U$;  then eliminate $\tensor{M}$ using eq. (\ref{eq:defU}):
 \begin{eqnarray}
\frac{\partial\tensor{U}}{\partial t}
&=&
-\vec{\nabla} \cdot {\cal J}_{U}+\tensor{V}-\tensor{\Omega}\tensor{U}
\nonumber \\
&&-\tensor{U}\tensor{\Omega}^{t}+{\cal O}\left(\tensor{\Omega}\tensor{U}^{2}\right)-\tensor{P}.
\label{kine}
\end{eqnarray}

The term $-\tensor{P}$ appears on the r.h.s. of eq. (\ref{kine})
thanks to the factor $-1/2$ in eq. (\ref{eq:defP}). Physically, we can track it across
section \ref{upperconvectedderivative}, back to the factor 2 in
 eq. (\ref{inverti}), thus in  eq. (\ref{eq:defU}).

The higher order terms ${\cal O}(\tensor{\Omega}\tensor{U}^{2})$
     are often negligible in a plastic material such as considered here, where
the elastic internal strain is seldom much larger than unity.

On the other hand, the advection term  ${\cal J}_{U}$
and the rotation term
$\tensor{\Omega}\tensor{U} + \tensor{U}\tensor{\Omega}^{t}$
have symmetries which are different from that of
$\tensor{V}$ and $\tensor{P}$. They  may thus not necessarily be negligible.
They can be regrouped using the total {\it corotationnal} ("Jaumann" \cite{Pleiner2004}) objective derivative:\begin{equation}
\frac{{\cal D}\tensor{U}}{{\cal D}t}=\frac{\partial\tensor{U}}{\partial t}+\vec{\nabla} \cdot {\cal J}_{U}+\tensor{\Omega}\tensor{U}-\tensor{U}\tensor{\Omega},
\label{eq:corotationnal}
\end{equation}
where we recall that $\tensor{\Omega}^{t}= - \tensor{\Omega}$.
 We thus approximately obtain eq. (\ref{eq:Vshared}).
 This result  is close to the objective derivative  
 for the logarithmic Hencky strain (up to a small correction on the rotation rate),
 see  \cite{Xiao1998,Mora2004}.


\end{document}